\documentclass[journal]{IEEEtran}

\usepackage{setspace}
\usepackage{lettrine}

\usepackage{graphics,
	psfrag,
	epsfig,
	amsthm,
	cite,
	amssymb,
	url,
	dsfont,
	algorithm,
	algorithmic,
	balance,
	enumerate,
	color,
	setspace,
	subcaption,
	caption,
	multirow
}
\usepackage{amsmath}

\usepackage{epstopdf}

\newtheorem{theorem}{Theorem}
\newtheorem{corollary}{Corollary}

\newcommand{\eref}[1]{(\ref{#1})}
\newcommand{\sref}[1]{Section~\ref{#1}}

\newcommand{\fref}[1]{Figure~\ref{#1}}

\newcommand{\cref}[1]{Constraint~\ref{#1}}
\newcommand{\thref}[1]{Theorem~\ref{#1}}

\newcommand{\algref}[1]{Algorithm~\ref{#1}}

\newcommand{\ignore}[1]{}

\IEEEoverridecommandlockouts
\newcommand\blfootnote[1]{%
	\begingroup
	\renewcommand\thefootnote{}\footnote{#1}%
	\addtocounter{footnote}{-1}%
	\endgroup
}

\usepackage{etex}
\IEEEoverridecommandlockouts
\usepackage{amsmath}
\usepackage{graphicx}
\usepackage{amsthm}
\usepackage{amssymb}
\usepackage{epstopdf}
\usepackage{epsfig}
\usepackage{listings}
\usepackage{mathtools,cuted}
\usepackage{multirow}
\usepackage{hhline}
\usepackage{tikz}
\usepackage{caption}
\usetikzlibrary{backgrounds,automata}
\usepackage{dblfloatfix} 
\usepackage{float}
\usepackage{placeins}
\usepackage{multicol}
\usepackage{lipsum}
\usepackage{cuted}
\usepackage{algorithm}
\usepackage{algorithmic}
\usepackage{tikz}
\usepackage{multicol, blindtext}
\usepackage{mathrsfs}

\graphicspath{{./figs/}}

\begin{document}

\title{\vspace{-.5cm}Manifold Optimization for High-Accuracy Spatial Location Estimation Using Ultrasound Waves}

\author{
   \IEEEauthorblockN{Mohammed H. AlSharif, \textit{Student Member, IEEE},  Ahmed Douik, \textit{Student Member, IEEE}, \\ Mohanad Ahmed, Tareq Y. Al-Naffouri, \textit{Senior Member, IEEE}, and Babak Hassibi, \textit{Member, IEEE}\vspace{-.8cm}}

\thanks {
	Mohammed Alsharif, Mohanad Ahmed, and Tareq Al-Naffouri are with the Division of Computer, Electrical and Mathematical Sciences, and Engineering, King Abdullah University of Science and Technology, Thuwal 23955-6900, Saudi Arabia (e-mail: \{mohammed.alsharif,tareq.alnaffouri\}@kaust.edu.sa, m.a.m.elhassan@gmail.com).
	
Ahmed Douik and Babak Hassibi are with the Department of Electrical Engineering, California Institute of Technology, Pasadena, CA 91125 USA (e-mail: \{ahmed.douik,hassibi\}@caltech.edu).
}
}

\maketitle
\blfootnote{\copyright 2021 IEEE.  Personal use of this material is permitted.  Permission from IEEE must be obtained for all other uses, in any current or future media, including reprinting/republishing this material for advertising or promotional purposes, creating new collective works, for resale or redistribution to servers or lists, or reuse of any copyrighted component of this work in other works.}
\begin{abstract}
This paper reports the design of a high-accuracy spatial location estimation method using ultrasound waves by exploiting the fixed geometry of the transmitters. Assuming an isosceles triangle antenna configuration, {where three antennas are placed as the vertices of an isosceles triangle}, the spatial location problem can be formulated as a non-convex optimization problem whose interior is shown to admit a Riemannian manifold structure. Our investigation of the geometry of the newly introduced manifold (i.e., the manifold of all isosceles triangles in $\mathbb{R}^3$) enables the design of highly efficient optimization algorithms. Simulations are presented to compare the performance of the proposed approach with popular methods from the literature. The results suggest that the proposed Riemannian-based methods outperform the state-of-the-art methods. Furthermore, the proposed Riemannian methods require much less computation time compared to popular generic non-convex approaches.
\end{abstract}

\begin{IEEEkeywords}
Spatial location estimation, ultrasound waves, fixed transmitters geometry, Riemannian manifold optimization.
\end{IEEEkeywords}

\section{Introduction} \label{sec:int}

\lettrine[lines=2]{W}{ith} the rapidly increasing number of smartphones and the proliferation of the Internet of Things (IoT), location-based services have attracted increased interest in the last decade \cite{davidson2016survey}. These services range from outdoor localization, e.g., for navigation purposes, to accurate indoor pinpointing for applications such as robot steering, surveillance, video gaming, and virtual reality \cite{gu2009survey}. While outdoor localization is universally solved by the Global Navigation Satellite System (GNSS), such a system is not feasible indoors. Thus, indoor localization systems have been implemented using various competing technologies, including ultrasound waves \cite{borriello2000location}, radio-frequency \cite{whitehouse2007practical}, infrared radiation \cite{yuzbacsiouglu2005improved}, and laser signals \cite{amann2001laser}.

Light-based localization systems, i.e., radio, infrared, and laser signals, suffer from low accuracy or high hardware costs. Indeed, due to the high speed of light, without precise and costly synchronization, small timing errors result in significant localization errors \cite{narayanan2000doppler}. As a result, localization systems based on WiFi or Bluetooth technologies have low accuracy and require pre-calibration \cite{cypriani2009open}. Similarly, while radio-based approaches utilizing the time of flight\footnote{The time of flight is the time required for the signal to travel from the transmitter to the receiver.} (ToF) estimation do not require pre-calibration, these systems depend on exact synchronization. Finally, laser and infrared-based localization devices are complicated and expensive to build and maintain\cite{rasshofer2005automotive}. This paper considers ultrasound-based localization methods \cite{ijaz2013indoor} due to their low cost and high accuracy, which is enabled by the relatively low speed of sound \cite{kushwaha2005sensor}.

Besides the effects of the employed technology, the accuracy of indoor localization systems primarily depends on the optimization algorithms utilized in the design of those systems, e.g., see \cite{seco2009survey,zafari2019survey}, and references therein. For example, a simple approach consists of estimating the received signal strength (RSS). While popular, RSS-based methods suffer from poor localization accuracy due to multipath fading and temporal dynamics \cite{yang2013rssi}. Alternatively, the Angle of Arrival (AoA) can be exploited to design high-accuracy systems for close-range location estimation. However, its performance significantly degrades as the distance between the transmitter and the receiver increases. This deterioration in accuracy is a consequence of the fact that a tiny error in the estimated angle results in a massive failure in the estimated position \cite{kumar2014accurate}.

The previously mentioned ToF-based approach represents an attractive alternative due to its simplicity. Still, a small perturbation in the estimated ToF can result in a significant deviation in the expected location, especially under an unfavorable geometry \cite{langley1999dilution}. To circumvent the aforementioned limitation, this manuscript uses multiple transmitters and considers exploiting their geometry in the estimation process. The resulting transmitter diversity not only significantly improves the accuracy of the estimated location but also provides the $3$D orientation of the device.

In particular, this paper considers a target with three transmitters that are placed on an {isosceles} triangle and utilizes a set of four receivers, known as beacons, to accurately estimate the $3$D location and orientation of the target. The positions of the beacons are assumed to be identified correctly. All of the distances between the transmitters and receivers are accurately estimated using our ranging algorithm \cite{alsharif2017zadoff}. {In conventional localization methods, these distances are fed into a classical non-linear least squares solver, such as the Gauss-Newton algorithm \cite{wright1999numerical}, to obtain the $3$D locations of the transmitters. In this paper, the measured distances are exploited through novel non-convex Riemannian-based optimization algorithms to obtain more accurate location estimates of the three transmitters.}

The main contribution of this paper is to provide a novel and highly accurate spatial location estimation method. To that end, the geometry of the transmitters is integrated into the location estimation process by formulating the problem as a non-convex optimization. Subsequently, the set of feasible solutions is shown to admit a Riemannian manifold structure, which enables the underlying optimization problem to be rigorously solved. To the best of the authors' knowledge, the {isosceles} triangle manifold has not been introduced nor studied in the literature. Hence, this manuscript characterizes its geometry in order to design Riemannian optimization algorithms for the ultrasound spatial location estimation problem. The efficiency of the proposed method is validated through extensive simulations and comparisons against the-state-of-the-art algorithms in the literature. The numerical results suggest that the inclusion of the fixed {isosceles} triangle geometry of the transmitters, as non-linear constraints in the optimization problem, significantly improves the quality of the location estimates. In fact, we confirm these results by deriving the Cram\'er-Rao bound (CRB) and the constrained CRB for our localization setup.  Furthermore, the proposed Riemannian-based method offers a clear advantage in terms of reduced complexity{, evaluated through the computational time,} compared to popular generic non-convex approaches.

The rest of this manuscript is organized as follows. \sref{sec:rel} cites and discusses the related work. In \sref{sec:sys}, the {isosceles} triangle geometry of the transmitters is exploited to formulate the ultrasound spatial location estimation problem as a non-convex optimization. A brief overview of Riemannian optimization methods on embedded manifolds is provided in \sref{sec:opt}. The {isosceles} triangle manifold is introduced and investigated in \sref{sec:the}. Given the manifold geometry, \sref{sec:hig} describes the design of efficient algorithms for the location estimation problem. {\sref{sec:ccrb} derives the Cram\'er-Rao bound (CRB) and the constrained CRB for our localization setup.} Finally, \sref{sec:sim} presents the simulation results and comparisons with other methods from the literature, before concluding in \sref{sec:con}.

\section{Related Work} \label{sec:rel}

A significant portion of indoor localization works described in the literature utilize an array of receivers or transmitters to determine the location and orientation of a target. For example, in \cite{ward1997new}, the authors designed the Active Bat system, which estimates the position and orientation of an array of ultrasound transmitters based on the ToF estimation. The reference suggests performing nonlinear regression combined with a least-squares solver to obtain the position of the target. The Active Bat system exhibits an accuracy of $3$ cm in a 3D space \cite{zafari2019survey}. This high accuracy is achieved at the cost of requiring many well-placed beacons. Similarly, the Cricket system is introduced in \cite{priyantha2000cricket}. This system consists of an array of ultrasound receivers that estimates the time of arrival (ToA) and angle of arrival (AoA) simultaneously. The ToA and AoA are fed to a nonlinear least squares solver to obtain the location and orientation of the target. In contrast to the Active Bat, the Cricket system does not require synchronization between the transmitter and receiver. However, the Cricket system only achieves an accuracy of 10 cm \cite{Jie2015Push}. In \cite{fukuju2003dolphin}, DOLPHIN was introduced as a system to localize synchronized nodes in a typical indoor environment. DOLPHIN is similar to the Active Bat and Cricket systems, except that it only requires few pre-configured reference nodes. Experimental evaluation of DOLPHIN reported an accuracy of 15 cm in a 2D space, with a possible degradation in accuracy observed as the algorithm was extended to a 3D space \cite{fukuju2003dolphin}. 
 
Combining the ToA and AoA as in \cite{priyantha2000cricket} results in better performance, which explains the wide adoption of these techniques in the literature. For example, Saad et al. \cite{saad2012high} extended the method to mobile devices in a system that utilizes an array of three receivers. However, instead of relying on a nonlinear least-squares solution, the authors estimate the position of the target through a classical trilateration algorithm. The hybrid method in \cite{saad2012high} can achieve a localization accuracy of $7.8$ cm \cite{7140728}, at the expense of the higher computational cost required to simultaneously estimate the ToF and AoA. Along the same lines, reference \cite{gonzalez2009high} considers an array of eight receivers whose locations are estimated by a Taylor series trilateration method \cite{foy1976position}. Finally, the position and orientation of the mobile device are obtained by averaging the AoA and location estimates for the eight receivers. While the experimental evaluation in \cite{gonzalez2009high} achieved  good location and orientation accuracies, the distances between the receivers must be less than $3.43$ mm, which is impossible to achieve with conventional ultrasound sensors. The classical trilateration algorithm has been extended in \cite{hazas2002novel} and \cite{hazas2006broadband} into a multilateration algorithm for a broadband localization system, with multiple ultrasound transmitters and receivers, as an improvement to the Cricket system. Nevertheless, this enhancement demands costly digital signal processing (DSP) techniques. 

The authors in \cite{kim2012dynamic} consider the problem of determining the position of a moving robot in a system comprised of an equilateral ultrasound receiver array with a set of transmitters at known locations. The problem is solved by extending the Kalman filter to incorporate a dynamic distance estimation method. The maximum error in the robot position in \cite{kim2012dynamic} is $25.7$ cm, which is quite large for many location-based applications. 

To the best of the authors' knowledge, all previously reported indoor localization systems do not impose the geometry of the array of receivers as a constraint when solving for the position of the target. Exploiting the fixed geometry of the transmitters or receivers array when formulating the optimization problem is expected to improve the localization estimation accuracy, as long as the resulting non-convex problem can be solved efficiently. To this end, the rest of this manuscript formulates the location determination problem as a non-convex program in which the constraints highlight the geometry of the transmitters. Subsequently, an efficient Riemannian-based optimization algorithm is designed by studying the geometry of the manifold, which is derived from the set of all feasible solutions.

\section{Location Estimation using Ultrasound Waves} \label{sec:sys}

\subsection{System Model and Parameters}

\begin{figure}[t]
\centering
\includegraphics[width=1\linewidth]{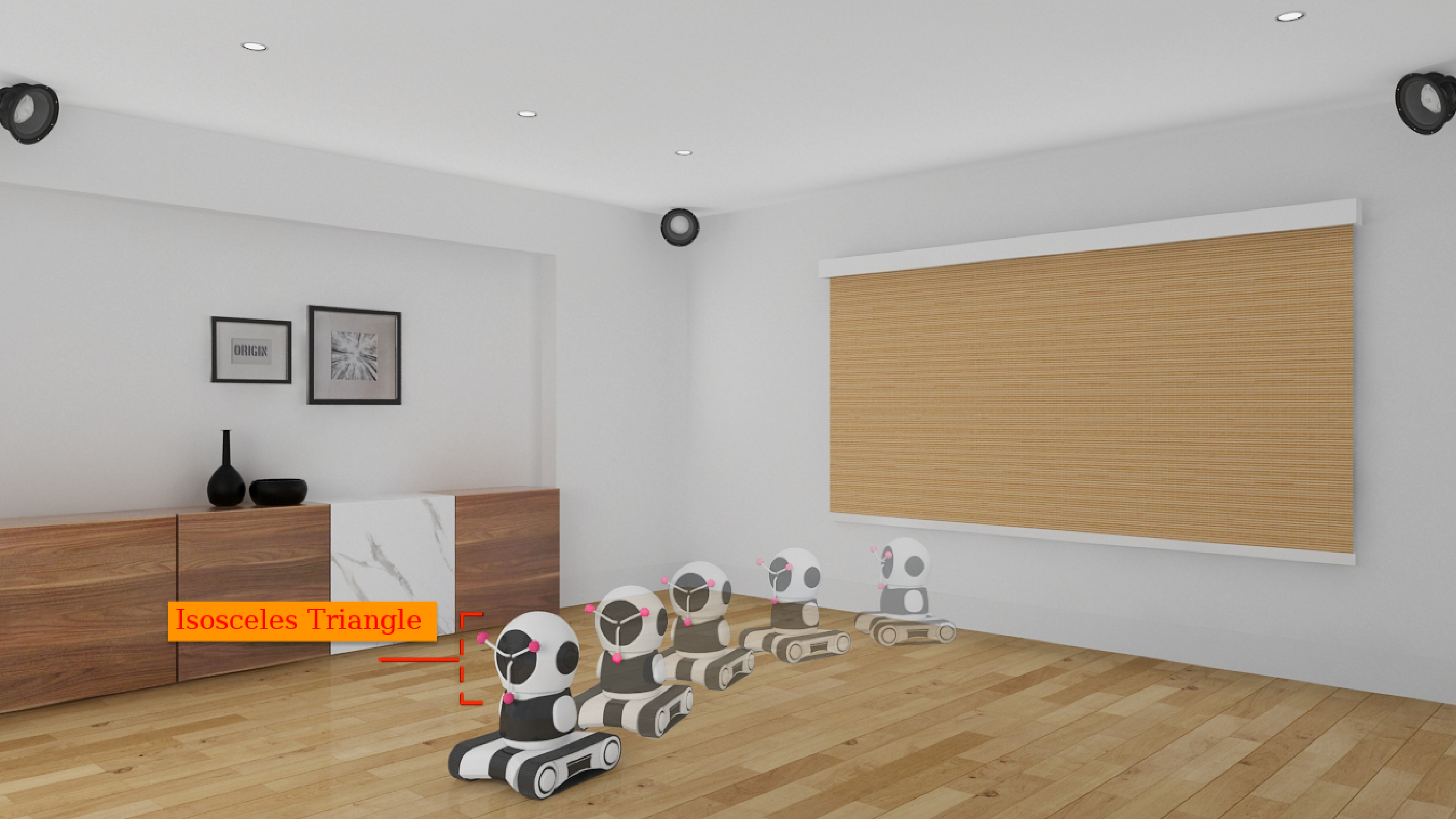}
\caption{A localization system consisting of three ultrasound transmitters and four receivers (beacons). The position of the target (robot) with respect to the transmitters is known.} \label{prob_set}
\end{figure}

This paper considers a localization system consisting of three ultrasound transmitters and four receivers, also known as beacons, whose exact positions are known. The three transmitters form an {isosceles} triangle of {base} length $d$, as shown in \fref{prob_set}. The position of the target of interest is known with respect to the transmitters, e.g., the target can be placed in the centroid of the triangle. Therefore, estimating the $3$D locations of the three transmitters accurately provides the $3$D location and orientation of the target. {We would like to mention that swapping the roles of the transmitters and receivers --- i.e., a three-receiver array becomes the target and the four transmitters become the beacons --- doesn't require any changes in the proposed algorithm.}

Let the $3$D location of the $i$-th transmitter be $\mathbf{x}_i \in \mathbb{R}^3$. Likewise, let $\mathbf{b}_j \in \mathbb{R}^3$ denote the position of the $j$-th beacon. These positions are grouped in a matrix $\mathbf{A} \in \mathbb{R}^{4 \times 3}$ such that the $j$-th row of the matrix corresponds to the location of the $j$-th beacon, i.e., $\mathbf{a}^T_j = \mathbf{b}^T_j$ where the notation $\mathbf{z}^T$ refers to the transpose of the vector $\mathbf{z}$.

{
Let the received signal from the $i^{\textit{th}}$ transmitter at the $j^{\textit{th}}$ receiver be given by \cite{5466526}
\begin{equation}
\zeta_{ij}(t) = \psi_{ij} s_i(t-\kappa_{ij})+n_{ij} (t), \; i=1,2,...,M, \; j=1,2,...,N,
\label{ddd}
\end{equation}
where $\psi_{ij}$ is an attenuation factor incurred by propagation, $s_i (t)$ is the passband transmitted signal from the $i^{\textit{th}}$ transmitter, $\kappa_{ij}$ is the time of flight (ToF) from the $i^{\textit{th}}$ transmitter to the $j^{\textit{th}}$ receiver, $M$ is the number of transmitters, $N$ is the number of receivers, and $n_{ij}(t)$ is an additive Gaussian noise with zero mean and variance $\sigma_{ij}^2$. We obtain the distances between the transmitters and receivers by multiplying the ToF $\kappa_{ij}$ by the speed of sound.
}
The range $r_{ij}$ (i.e., distance) from the $i$-th transmitter to the $j$-th beacon is estimated from the measurements of the received signal {using \cite{alsharif2017zadoff} and \cite{alsharif2021range}}. Thus, the distance can be expressed using the vectors $\mathbf{x}_i$ and $\mathbf{b}_j$ as follows
\begin{align} \label{eq:-1}
r_{ij} = ||\mathbf{x}_i-\mathbf{b}_j||_2= \sqrt{(\mathbf{x}_i-\mathbf{b}_j)^T(\mathbf{x}_i-\mathbf{b}_j)}.
\end{align}
For ease of notation, the positions, distances, and measurements are collected in vectors as follows
{\begin{align*}
 ||\mathbf{x}_i||_2^2 \mathbf{1}_{4} = \begin{pmatrix}
||\mathbf{x}_i||_2^2 \\
||\mathbf{x}_i||_2^2 \\
||\mathbf{x}_i||_2^2 \\
||\mathbf{x}_i||_2^2 
\end{pmatrix}, \mathbf{b}^2 = \begin{pmatrix}
||\mathbf{b}_1||_2^2 \\
||\mathbf{b}_2||_2^2 \\
||\mathbf{b}_3||_2^2 \\
||\mathbf{b}_4||_2^2 
\end{pmatrix}, \mathbf{r}_{i}^2 = \begin{pmatrix}
r_{i1}^2 \\
r_{i2}^2 \\
r_{i3}^2 \\
r_{i4}^2
\end{pmatrix},
\end{align*}}
where $\mathbf{1}_{4}$ is the all-ones vector of dimension $4$. The transformed measurement vector can be defined as $\mathbf{y}_i = \cfrac{1}{2}(\mathbf{b}^2-\mathbf{r}^2_i)$. Using the expression of $r_{ij}$ in \eref{eq:-1}, it can easily be concluded that 
{\begin{align} \label{eq:-2}
\mathbf{A}\mathbf{x}_i - \frac{1}{2}||\mathbf{x}_i||_2^2 \mathbf{1}_{4} = \mathbf{y}_i \in \mathbb{R}^{4 \times 1}.
\end{align}}

\subsection{Location Estimation Problem Formulation}

This manuscript assumes that all the distances $r_{ij}$, are affected by a normally distributed noise. Therefore, a reasonable objective function is to consider the $\ell_2$ loss between the measurement and the model \eref{eq:-2}. In other words, the paper considers the following objective function
{\begin{align*}
\sum_{i=1}^3 ||\mathbf{A}\mathbf{x}_i - \frac{1}{2}||\mathbf{x}_i||_2^2 \mathbf{1}_{4} - \mathbf{y}_i||_2^2.
\end{align*}}

The choice of the loss function to be used depends on the assumptions on the system model. While this manuscript focuses on the $\ell_2$ loss, the results are more generic and can be applied to any smooth loss function, as explained at the end of this section. To incorporate the fixed geometry of the transmitters, the spatial location estimation problem using ultrasound waves can be formulated as
{\begin{subequations}
\label{eq:1}
\begin{align}
\min_{\mathbf{x}_1,\mathbf{x}_2,\mathbf{x}_3 \in \mathbb{R}^{3}} \ & \sum_{i=1}^3 ||\mathbf{A}\mathbf{x}_i - \frac{1}{2}||\mathbf{x}_i||_2^2 \mathbf{1}_{4} - \mathbf{y}_i||_2^2 \label{eq:2} \\
\label{eq:3} {\rm s.t.\ } & (\mathbf{x}_1-\mathbf{x}_2)^T(\mathbf{x}_2-\mathbf{x}_3) = - d^2 \cos(\frac{\pi}{3})\\
\label{eq:4} & (\mathbf{x}_1-\mathbf{x}_3)^T(\mathbf{x}_2-\mathbf{x}_3) = d^2 \cos(\frac{\pi}{3}),
\end{align}
\end{subequations}}
{wherein the constraints \eref{eq:3} and \eref{eq:4} fix the length of the base $||\mathbf{x}_2 - \mathbf{x}_3||_2 = d$ and guarantee that both side legs of the triangle are of equal length $||\mathbf{x}_1 - \mathbf{x}_2||_2 = ||\mathbf{x}_1 - \mathbf{x}_3||_2$.} {Moreover, we restrict the manifold to not include a set of measure zero defined by $\mathbf{x}_1^T(\mathbf{x}_2-\mathbf{x}_3) = 0$. The reason for this is seen in section V. While this restriction seems to exclude some points that need to be localized, the set of feasible points include all {isosceles} triangles in the localization problem as will be explained in section V.}

Despite the convexity of the objective function in \eref{eq:2}, the optimization problem is non-convex due to the quadratic nature of the constraints. Indeed, each feasible solution $\mathbf{x}_1,\mathbf{x}_2,\mathbf{x}_3 \in \mathbb{R}^{3}$ to \eref{eq:1} belongs to a set, named the \emph{{isosceles} triangle manifold}, which is defined as follows:
{\begin{align*}
\mathcal{M} = \Big\{\{\mathbf{x}_i\}_{i=1}^3 \in \mathbb{R}^{3} \ \Big| (\mathbf{x}_1-\mathbf{x}_2)^T(\mathbf{x}_2-\mathbf{x}_3) &= -d^2 \cos(\frac{\pi}{3}) \\
(\mathbf{x}_1-\mathbf{x}_3)^T(\mathbf{x}_2-\mathbf{x}_3) &= d^2 \cos(\frac{\pi}{3})  \\
\mathbf{x}_1^T(\mathbf{x}_2-\mathbf{x}_3) &\neq 0\Big\}.
\end{align*}}

The study of the geometry of this newly introduced manifold allows us to take advantage of Riemannian optimization methods to efficiently solve the location estimation problem. Furthermore, instead of directly solving the optimization problem \eref{eq:1}, this manuscript suggests solving its generalization. In particular, let $\{\mathbf{x}_{i}\}_{i=1}^3$ be $3$-dimensional vectors in $\mathbb{R}^3$ and consider a smooth function $f: \left(\mathbb{R}^{3}\right)^3 \longrightarrow \mathbb{R}$, that may or may not be convex, the rest of this manuscript solves the optimization problem
\begin{subequations}
\label{eq:5}
\begin{align}
\min_{\mathbf{x}_1,\mathbf{x}_2,\mathbf{x}_3 \in \mathbb{R}^{3}} \ &f(\mathbf{x}_1,\mathbf{x}_2,\mathbf{x}_3) \label{eq:6} \\
\label{eq:7} {\rm s.t.\ } & (\mathbf{x}_1,\mathbf{x}_2,\mathbf{x}_3) \in \mathcal{M}.
\end{align}
\end{subequations}

\section{Optimization on Riemannian Manifolds} \label{sec:opt}

This manuscript uses Riemannian manifold optimization to solve the location estimation problem described in the previous section. {In this paper, we consider submanifolds of Euclidean spaces equipped with the corresponding Euclidean metric. Since this section succinctly introduces all of the required ingredients by presenting an overview of Riemannian optimization methods over matrix manifolds{\cite{absil2009optimization}}, no prior knowledge of differential geometry or Riemannian manifold is required from the reader.} More specifically, while Subsection \ref{sub:1} recalls all necessary manifold terminology, definitions, and notations, Subsection \ref{sub:2} illustrates the design of a Riemannian optimization algorithm and recalls relevant convergence results. {For more details on Riemannian optimization we refer the reader to \cite{absil2009optimization}, \cite{smith1994optimization}, and \cite{liu2019simple}.}

\subsection{Manifold Terminology, Definitions, and Notations} \label{sub:1}

A matrix manifold $\mathcal{M}$ embedded in the Euclidean space of matrices $\mathcal{E} = \mathbb{R}^{n \times m}$ is a subset of $\mathcal{E}$ that is in bijection with an open space $\mathcal{E}^* \subseteq \mathcal{E}$. At each point $\mathbf{X} \in \mathcal{M}$, the manifold can be locally approximated by a {$\rho$-dimensional} linear space known as the \emph{tangent space} and denoted by $\mathcal{T}_{\mathbf{X}} \mathcal{M}$. Tangent spaces play a primordial role in the design of optimization algorithms in the same fashion as derivatives are crucial to approximate functions. Indeed, the tangent space $\mathcal{T}_{\mathbf{X}} \mathcal{M}$ at $\mathbf{X} \in \mathcal{M}$ can be seen as a local linearization of the manifold $\mathcal{M}$ around $\mathbf{X}$. Furthermore, the dimension {$\rho$} of $\mathcal{T}_{\mathbf{X}} \mathcal{M}$ is said to be the dimension of the manifold  $\mathcal{M}$ as it represents its degrees of freedom. 

In order to perform optimization on a manifold $\mathcal{M}$, one needs the notion of {length that applies to tangent vectors. This is accomplished by} endowing each tangent space $\mathcal{T}_{\mathbf{X}} \mathcal{M}$ by an inner product $\langle .,. \rangle_{\mathbf{X}}$ that is smoothly varying for $\mathbf{X} \in \mathcal{M}$. {This smoothly varying inner product is known as the Riemannian metric, and the manifold $\mathcal{M}$ is called a Riemannian manifold}{\cite{absil2009optimization}}. While multiple Riemannian metrics exist for a manifold $\mathcal{M}$, a canonical choice for embedded matrix manifolds is the canonical inner product of the matrix space. A particular property of induced Riemannian metrics from the Euclidean space of matrices is that they do not depend on the origin of the tangent space, i.e., $\langle .,. \rangle_{\mathbf{X}} = \langle .,. \rangle$ for any $\mathbf{X} \in \mathcal{M}$. The rest of this manuscript assumes that the canonical Riemannian metric is utilized as it allows us to simplify the expressions of the geometry of the manifold, e.g., gradient, covariant derivative, and Hessian. 

Let $f: \mathcal{M} \longrightarrow \mathbb{R}$ be a smooth function from the matrix manifold $\mathcal{M}$ to $\mathbb{R}$ and let $\mathbf{X} \in \mathcal{M}$ be a point on the manifold. Consider a tangent vector $\xi_{\mathbf{X}} \in \mathcal{T}_{\mathbf{X}} \mathcal{M}$, the derivative of $f(\mathbf{X})$ in the direction $\xi_{\mathbf{X}}$, denoted by $\text{D}(f(\mathbf{X}))[\xi_{\mathbf{X}}]$, is defined as
\begin{align*}
\text{D}(f(\mathbf{X}))[\xi_{\mathbf{X}}] = \lim_{h \rightarrow 0} \cfrac{f(\mathbf{X}+h \xi_{\mathbf{X}}) - f(\mathbf{X})}{h}. 
\end{align*}

One can note that the definition of directional derivatives above is closely related to the one for Euclidean spaces, with the exception that only tangent vectors $\xi_{\mathbf{X}} \in \mathcal{T}_{\mathbf{X}} \mathcal{M}$ are permitted as perturbations. Indeed, while the expression $f(\mathbf{X}+ \mathbf{Y})$ is well defined for a Euclidean space $\mathcal{E}$ thanks to the fact that $\mathbf{X}+ \mathbf{Y} \in \mathcal{E}$ for all vectors $\mathbf{X}, \mathbf{Y} \in \mathcal{E}$, for a Riemannian manifold $\mathcal{M}$, the expression $f(\mathbf{X}+h \xi_{\mathbf{X}})$ is only valid for a tangent vector $\xi_{\mathbf{X}} \in \mathcal{T}_{\mathbf{X}} \mathcal{M}$ and a small perturbation $h \ll 1$. In fact, as stated earlier, the tangent space $\mathcal{T}_{\mathbf{X}}\mathcal{M}$ is only a linear approximation of the manifold locally around $\mathbf{X}$. The indefinite directional derivative of $f$ at $\mathbf{X}$ is defined as the operator $\text{D}(f(\mathbf{X})): \mathcal{T}_{\mathbf{X}}\mathcal{M} \longrightarrow \mathbb{R}$ which associates to each $\xi_{\mathbf{X}}$ the directional derivative $\text{D}(f(\mathbf{X}))[\xi_{\mathbf{X}}]$.

\subsection{Optimization Over Matrix Manifolds} \label{sub:2}

Riemannian optimization is an extension of unconstrained iterative optimization methods from Euclidean spaces to Riemannian manifolds. In other words, Riemannian optimization methods can be viewed as unconstrained optimization over a constrained set, i.e., a manifold. Recall that unconstrained Euclidean optimization starts with an initial guess $\mathbf{X} \in \mathcal{E}$ and at each iteration finds a descent direction $\xi$ and a step size $\alpha$ to update the point through $\mathbf{X} = \mathbf{X} + \alpha \xi$. The aforementioned steps of unconstrained Euclidean optimization are summarized in \algref{alg1}.  

\begin{algorithm}[t!]
\begin{algorithmic}[1]
\STATE Initialize $\mathbf{X} \in \mathcal{E}$.
\WHILE {$||\nabla_{\mathbf{X}} f|| \neq 0$}
\STATE Find a descent direction $\xi \in \mathcal{E}$ using $\nabla_{\mathbf{X}} f$ and/or $\nabla^2_{\mathbf{X}} f$
\STATE Compute a step size $\alpha$.
\STATE update $\mathbf{X} = \mathbf{X} + \alpha \xi$.
\ENDWHILE
\end{algorithmic}
\caption{Template of Unconstrained Optimization}
\label{alg1}
\end{algorithm}

As stated earlier, Riemannian optimization extends \algref{alg1} to Riemannian manifolds. While unconstrained optimization initialize with any $\mathbf{X} \in \mathcal{E}$, Riemannian optimization require a feasible point $\mathbf{X} \in \mathcal{M}$. Afterwards, the curvature of the manifold is approximated locally around $\mathbf{X}$ by a linear {$\rho$-dimensional} space through the computation of the tangent space $\mathcal{T}_{\mathbf{X}} \mathcal{M}$. Since the tangent space is linear, one can find a descent direction in that tangent space using the same technique as for unconstrained optimization. However, this requires the introduction of the Riemannian gradient $\overline{\nabla}_{\mathbf{X}} f$ and Hessian $\overline{\nabla}^2_{\mathbf{X}} f$ as the Euclidean gradient ${\nabla}_{\mathbf{X}} f$ and Hessian ${\nabla}^2_{\mathbf{X}} f$ are defined on the original high-dimensional space $\mathcal{E}$ and not exclusively on $\mathcal{T}_{\mathbf{X}} \mathcal{M}$.

Thanks to the use of the canonical Riemannian metric induced from the embedding space, the Riemannian gradient $\overline{\nabla}_{\mathbf{X}} f$ can be obtained by simply projecting the Euclidean gradient ${\nabla}_{\mathbf{X}} f$ onto the tangent space $\mathcal{T}_{\mathbf{X}} \mathcal{M}$, i.e, $\overline{\nabla}_{\mathbf{X}} f = \Pi_{\mathbf{X}}({\nabla}_{\mathbf{X}} f)$. Such an orthogonal projection $\Pi_{\mathbf{X}}$, {i.e. orthogonal according to the Riemannian metric}, onto the tangent space $\mathcal{T}_{\mathbf{X}} \mathcal{M}$ is well defined for any $\mathbf{X} \in \mathcal{M}$ due to the linear structure of the tangent space. Similarly, the definition of the Riemannian Hessian $\overline{\nabla}^2 _{\mathbf{X}} f$ follows that of the Euclidean Hessian, which is obtained as the directional derivative of the gradient, i.e., ${\nabla}^2_{\mathbf{X}} f[\xi_{\mathbf{X}}] = \text{D}({\nabla}_{\mathbf{X}} f)[\xi_{\mathbf{X}}]$. However, as the directional derivative of the Riemannian gradient may not be contained in the tangent space, the Riemannian Hessian requires projecting the directional derivative of the Riemannian gradient onto the tangent space, i.e.,
\begin{align} \label{eq:15}
\overline{\nabla}^2_{\mathbf{X}} f[\xi_{\mathbf{X}}] = \Pi_{\mathbf{X}}(\text{D}(\overline{\nabla}_{\mathbf{X}} f)[\xi_{\mathbf{X}}])
\end{align}

Finally, after we obtain a descent direction $\xi_{\mathbf{X}}$, and a step size $\alpha$ {depending on the employed algorithm}, Euclidean algorithms update the current position through the linear equation $\mathbf{X} = \mathbf{X} + \alpha \xi_{\mathbf{X}}$. While such an update is valid for linear spaces, it does not apply to Riemannian manifolds as it might result in a point outside the manifold. Therefore, one needs to ``project" such an update on the manifold while conserving its descent property. This is naturally accomplished by moving along geodesics, i.e., straight lines, on the manifold. However, finding the expression of geodesics can be difficult \cite{AbsMahSep2008} which motivates the use of their first order approximation known as \emph{retractions}. In particular, a retraction $\text{R}_{\mathbf{X}}: \mathcal{T}_{\mathbf{X}} \mathcal{M} \longrightarrow \mathcal{M}$ is a mapping from tangent vectors to the manifold such as $\text{R}_{\mathbf{X}}(\mathbf{0}_{\mathbf{X}}) = \mathbf{X}$ and $\cfrac{d\ \text{R}_{\mathbf{X}}(h \xi_{\mathbf{X}})}{d\ h}\Big|_{h=0} = \xi_{\mathbf{X}}$. Therefore, deriving a computationally efficient retraction is a critical and delicate step in designing Riemannian optimization algorithms.

The steps of Riemannian optimization methods are summarized in the template shown in \algref{alg2}. {Unlike the unconstrained optimization in \algref{alg1}, \algref{alg2} requires the point $\mathbf{X}$ to be on the manifold, and the descent direction to be on the tangent space.} In the rest of this manuscript, an instance of \algref{alg2} in which the descent direction only requires the gradient information is referred to as a first-order Riemannian optimization algorithm while the use of the Riemannian Hessian elevates the algorithm to second-order. For example, choosing $\xi_{\mathbf{X}} = - \cfrac{\overline{\nabla}_{\mathbf{X}} f}{||\overline{\nabla}_{\mathbf{X}} f||_{\mathbf{X}}}$ yields the celebrated steepest descent algorithm on Riemannian manifolds {\cite{absil2009optimization}}. Likewise, Newton's method on Riemannian manifolds, a second-order algorithm, is obtained by finding the tangent vector $\xi_{\mathbf{X}} \in \mathcal{T}_{\mathbf{X}} \mathcal{M}$ that satisfies the following Newton's equation $\overline{\nabla}^2_{\mathbf{X}} f[\xi_{\mathbf{X}}] = -\overline{\nabla}_{\mathbf{X}} f$.

\begin{algorithm}[t!]
\begin{algorithmic}[1]
\STATE Initialize ${\mathbf{X}} \in \mathcal{M}$.
\WHILE {$||\overline{\nabla}_{\mathbf{X}} f||_{{\mathbf{X}}} \neq 0$}
\STATE Find a descent direction $\xi_{\mathbf{X}} \in \mathcal{T}_{\mathbf{X}}\mathcal{M}$ using $\overline{\nabla}_{\mathbf{X}} f$ and/or $\overline{\nabla}^2_{\mathbf{X}} f$
\STATE Compute the step size $\alpha$ using backtracking.
\STATE Retract ${\mathbf{X}} = \text{R}_{\mathbf{X}}(\alpha \xi_{\mathbf{X}})$.
\ENDWHILE
\end{algorithmic}
\caption{Template of Riemannian Optimization}
\label{alg2}
\end{algorithm}

\section{The {Isosceles} Triangle Manifold} \label{sec:the}

This section investigates and characterizes {the geometry} of the {isosceles} triangle manifold, so that --- in \sref{sec:hig} --- optimization algorithms can be employed for the location estimation problem of interest in this paper. 
The first part of this section shows that the set is indeed a manifold and computes its tangent space and orthogonal projection. The second and third parts of the section compute the expression of the Riemannian gradient and Hessian and derive a computationally efficient retraction of the {isosceles} triangle manifold.

For conciseness and ease of notation, the variable $\mathbf{X} \in \mathbb{R}^{3 \times 3}$ is used in the rest of this manuscript as a shorthand notation for the three vectors $\mathbf{X} = [\mathbf{x}_1,\mathbf{x}_2,\mathbf{x}_3]$. Similarly, the tangent vector at $\mathbf{X} = [\mathbf{x}_1,\mathbf{x}_2,\mathbf{x}_3]$ is denoted by $\xi_{\mathbf{X}} = [\xi_{\mathbf{x}_1},\xi_{\mathbf{x}_2},\xi_{\mathbf{x}_3}]\in \mathbb{R}^{3 \times 3}$. 

\subsection{Manifold Geometry and Operators}

Recall that the {isosceles} triangle manifold is defined by
{\begin{align} 
\mathcal{M} = \Big\{\mathbf{X} \in \mathbb{R}^{3 \times 3} \ \Big| (\mathbf{x}_1-\mathbf{x}_2)^T(\mathbf{x}_2-\mathbf{x}_3) &= -d^2 \cos(\frac{\pi}{3}) \nonumber \\
\hspace{2.3cm} (\mathbf{x}_1-\mathbf{x}_3)^T(\mathbf{x}_2-\mathbf{x}_3) &= d^2 \cos(\frac{\pi}{3}) \nonumber \\
\mathbf{x}_1^T(\mathbf{x}_2-\mathbf{x}_3) &\neq 0 \Big\} \label{eq:8}.
\end{align}}

To show that the set defined in \eref{eq:8} is a well-defined manifold and to compute its tangent space, this section uses the implicit function theorem \cite{AbsMahSep2008}, which can be formulated as follows 
\begin{theorem} \label{th1}
Let $\mathcal{E}$ be a Euclidean space and let $g: \mathcal{E} \longrightarrow \mathcal{E}^\prime$ be a smooth and \emph{constant-rank} function from $\mathcal{E}$ to a linear space $\mathcal{E}^\prime$. Under these conditions, any level-set $\mathcal{M}$ of $g$ admits a manifold structure. Furthermore, the tangent space at $\mathbf{X} \in \mathcal{M}$ is given by
\begin{align*}
\mathcal{T}_{\mathbf{X}}\mathcal{M} = \text{Ker}(\text{D}(g(\mathbf{X}))).
\end{align*}
\end{theorem}
{First of all, we show that the set $\mathcal{M}$ defined in \eref{eq:8} admits a manifold structure. Therefore, we define a function $g: \mathbb{R}^{3 \times 3} \longrightarrow \mathbb{R}^{2}$ as}
\begin{align}\label{eq:10}
g(\mathbf{X}) = \begin{pmatrix}
(\mathbf{x}_1-\mathbf{x}_2)^T(\mathbf{x}_2-\mathbf{x}_3)+ d^2 \cos(\frac{\pi}{3}) \\
(\mathbf{x}_1-\mathbf{x}_3)^T(\mathbf{x}_2-\mathbf{x}_3)- d^2 \cos(\frac{\pi}{3})
\end{pmatrix}.
\end{align}
The function defined above only involves linear combinations and inner products of the vectors, which makes the function smooth as required by the implicit function theorem. In addition, the set of $2$-dimensional vectors $\mathbb{R}^{2}$ is a linear space as mandated by \thref{th1}. From the definition of the function $g$ and the set $\mathcal{M}$ in \eref{eq:8}, it is clear that $\mathcal{M}$ is a level-set of $g$ as it can be interpreted as the image of $\mathbf{0} \in \mathbb{R}^{2}$. 

{Lastly, according to \thref{th1}, to prove that $\mathcal{M}$ is a well-defined manifold we need to show} that $g$ is a constant-rank function, which can be accomplished by demonstrating that $\mathbf{0}$ is a regular value of $g$, i.e., the rank of each $\mathbf{X} \in g^{-1}(\mathbf{0}) = \mathcal{M}$ is equal to $\text{Dim}(\mathbb{R}^{2}) =2$ or equivalently that the indefinite directional derivative of $g$ at any $\mathbf{X} \in \mathcal{M}$ is a surjective map.
Let $\mathbf{X} \in \mathcal{M}$ and consider an arbitrary direction $\xi_{\mathbf{X}} \in \mathbb{R}^{3 \times 3}$, the directional derivative of $g$ at $\mathbf{X}$ in the direction $\xi_{\mathbf{X}}$ is 
\begin{align} \label{eq:11}
\text{D} &(g(\mathbf{X}))[\xi_{\mathbf{X}}] = \\
&\begin{pmatrix}
(\xi_{\mathbf{x}_1}-\xi_{\mathbf{x}_2})^T(\mathbf{x}_2-\mathbf{x}_3) + (\mathbf{x}_1-\mathbf{x}_2)^T(\xi_{\mathbf{x}_2}-\xi_{\mathbf{x}_3}) \\
(\xi_{\mathbf{x}_1}-\xi_{\mathbf{x}_3})^T(\mathbf{x}_2-\mathbf{x}_3) + (\mathbf{x}_1-\mathbf{x}_3)^T(\xi_{\mathbf{x}_2}-\xi_{\mathbf{x}_3}) 
\end{pmatrix}. \nonumber
\end{align}
Let $\begin{pmatrix} \alpha \\ \beta \end{pmatrix}$ be an arbitrary vector in $\mathbb{R}^2$. Finding $\xi_{\mathbf{X}} \in \mathbb{R}^{3 \times 3}$ such that $\text{D}( g(\mathbf{X}))[\xi_{\mathbf{X}}] = \begin{pmatrix} \alpha \\ \beta \end{pmatrix}$ amounts to solving the following linear system of equations
\begin{align*}
\begin{pmatrix} 
\mathbf{x}_2 - \mathbf{x}_3 & \mathbf{x}_2 - \mathbf{x}_3 \\ 
\mathbf{x}_1 + \mathbf{x}_3 - 2\mathbf{x}_2 & \mathbf{x}_1 - \mathbf{x}_3\\
\mathbf{x}_2 - \mathbf{x}_1 & 2\mathbf{x}_3 - \mathbf{x}_1 - \mathbf{x}_2
\end{pmatrix}^T
\begin{pmatrix} 
\xi_{\mathbf{x}_1} \\
\xi_{\mathbf{x}_2} \\
\xi_{\mathbf{x}_3} 
\end{pmatrix}
= \begin{pmatrix} \alpha \\ \beta \end{pmatrix}.
\end{align*}
The above linear system has a fat matrix of dimension $2 \times 3$ and a full rank of $2$. Indeed, assuming that the rank is equal to $1$ gives the equality $\mathbf{x}_2 = \mathbf{x}_3$, which is impossible for any $\mathbf{X} \in \mathcal{M}$. Therefore, it holds true that the map $\text{D} (g(\mathbf{X}))$ is surjective, which concludes that the function $g$ is a constant rank function. {Therefore, according to the results of \thref{th1}, we conclude that the {isosceles} triangle set $\mathcal{M}$ defined in \eref{eq:8} is a well-defined manifold of dimension $7$ embedded in the Euclidean space $\mathbb{R}^{3 \times 3}$. }

{After showing that the set $\mathcal{M}$ admits a manifold structure, we derive its tangent space. According to \thref{th1},} the tangent space is given by all directions $\xi_{\mathbf{X}} \in \mathbb{R}^{3 \times 3}$ that nullify $\text{D} (g(\mathbf{X}))[\xi_{\mathbf{X}}]$, which --- according to the expression given in \eref{eq:11} --- can be written as
\begin{align} \label{eq:9}
\mathcal{T}_{\mathbf{X}}&\mathcal{M} = \Big\{\xi_{\mathbf{X}} \in \mathbb{R}^{3 \times 3} \ \Big|  \\
&(\xi_{\mathbf{x}_1}-\xi_{\mathbf{x}_2})^T(\mathbf{x}_2-\mathbf{x}_3) + (\mathbf{x}_1-\mathbf{x}_2)^T(\xi_{\mathbf{x}_2}-\xi_{\mathbf{x}_3}) = 0 \nonumber \\
&(\xi_{\mathbf{x}_1}-\xi_{\mathbf{x}_3})^T(\mathbf{x}_2-\mathbf{x}_3) + (\mathbf{x}_1-\mathbf{x}_3)^T(\xi_{\mathbf{x}_2}-\xi_{\mathbf{x}_3}) = 0  \Big\}. \nonumber
\end{align}
As stated earlier, this manuscript considers the induced Riemannian metric from the canonical inner product $\langle\mathbf{X}, \mathbf{Y}\rangle = \text{Tr}(\mathbf{X}^T\mathbf{Y})$ in $\mathbb{R}^{3 \times 3}$. In other words, the induced Riemannian metric on the tangent space $\mathcal{T}_{\mathbf{X}}\mathcal{M}$ is obtained from the natural embedding of $\mathcal{M}$ in $\mathbb{R}^{3 \times 3}$, i.e.,
\begin{align*}
\langle \xi_{\mathbf{X}},\eta_{\mathbf{X}} \rangle_{\mathbf{X}} = \text{Tr}(\xi_{\mathbf{X}}^T\eta_{\mathbf{X}}) = \xi_{\mathbf{x}_1}^T \eta_{\mathbf{x}_1} + \xi_{\mathbf{x}_2}^T \eta_{\mathbf{x}_2} + \xi_{\mathbf{x}_3}^T \eta_{\mathbf{x}_3}.
\end{align*}

\subsection{Riemannian Gradient and Hessian}

{In this part, we derive the expression of the Riemannian gradient and Hessian.} According to \sref{sec:opt}, the Riemannian gradient $\overline{\nabla}_{\mathbf{X}}f$ can be expressed as the orthogonal projection, {denoted as $\Pi_{\mathbf{X}}: \mathbb{R}^{3 \times 3} \longrightarrow \mathcal{T}_{\mathbf{X}}\mathcal{M}$,} of the Euclidean gradient from the embedding space $\mathbb{R}^{3 \times 3}$ to the tangent space $\mathcal{T}_{\mathbf{X}}\mathcal{M}$. 

{In order to derive the orthogonal projection, we start by defining the orthogonal complement $\mathcal{T}^\perp_{\mathbf{X}}\mathcal{M}$ to the tangent space $\mathcal{T}_{\mathbf{X}}\mathcal{M}$ at $\mathbf{X} \in \mathcal{M}$ as}
\begin{align*}
&\mathcal{T}^\perp_{\mathbf{X}}\mathcal{M} = \left\{\eta_{\mathbf{X}} \in \mathbb{R}^{3 \times 3} \ \Big|  \eta_{\mathbf{X}} = \mathbf{X} \mathbf{U}_{\alpha}^{\beta}\right\}, 
\end{align*}
for some reals $\alpha$ and $\beta$ and the matrix $\mathbf{U}_{\alpha}^{\beta}$ being constructed as
\begin{align*}
\mathbf{U}_{\alpha}^{\beta} = \begin{pmatrix}
0 & \alpha+\beta & -\alpha-\beta \\
\alpha+\beta & -2\alpha & \alpha-\beta \\ 
-\alpha-\beta & \alpha-\beta & 2 \beta
\end{pmatrix}.
\end{align*}
Indeed, consider a tangent vector $\xi_{\mathbf{X}} \in \mathcal{T}_{\mathbf{X}}\mathcal{M}$ and a normal vector $\eta_{\mathbf{X}} \in \mathcal{T}^\perp_{\mathbf{X}}\mathcal{M}$. Their inner product after expansion is given by
\begin{align*}
& \langle \xi_{\mathbf{X}},\eta_{\mathbf{X}} \rangle_{\mathbf{X}} = \xi_{\mathbf{x}_1}^T \eta_{\mathbf{x}_1} + \xi_{\mathbf{x}_2}^T \eta_{\mathbf{x}_2} + \xi_{\mathbf{x}_3}^T \eta_{\mathbf{x}_3} \\
& \quad =\alpha \left[ (\xi_{\mathbf{x}_1}-\xi_{\mathbf{x}_2})^T(\mathbf{x}_2-\mathbf{x}_3) + (\mathbf{x}_1-\mathbf{x}_2)^T(\xi_{\mathbf{x}_2}-\xi_{\mathbf{x}_3}) \right] \nonumber \\
&\quad \ \ + \beta \left[ (\xi_{\mathbf{x}_1}-\xi_{\mathbf{x}_3})^T(\mathbf{x}_2-\mathbf{x}_3) + (\mathbf{x}_1-\mathbf{x}_3)^T(\xi_{\mathbf{x}_2}-\xi_{\mathbf{x}_3}) \right] \\
& \quad= 0.
\end{align*}
Combining the above equality with the fact that $\mathcal{T}^\perp_{\mathbf{X}}\mathcal{M}$ is a Euclidean space of dimension $2$ allows us to conclude that $\mathcal{T}^\perp_{\mathbf{X}}\mathcal{M}$ represents the complement of the tangent space.

{The next stage is to derive the orthogonal projection from the embedding space to the tangent space, by utilizing the complement to the tangent space.} Let $\mathbf{Z} \in \mathbb{R}^{3 \times 3}$ be a vector in the ambient space and $\mathbf{X} \in \mathcal{M}$ be a point on the manifold. The vector $\mathbf{Z}$ can be decomposed into a tangent part $\Pi_{\mathbf{X}}(\mathbf{Z})=\mathbf{Z}_{\mathbf{X}} \in \mathcal{T}_{\mathbf{X}}\mathcal{M}$ and an orthogonal part $\Pi^\perp_{\mathbf{X}}(\mathbf{Z})=\mathbf{Z}_{\mathbf{X}}^\perp \in \mathcal{T}_{\mathbf{X}}^\perp\mathcal{M}$. From the previous analysis of the orthogonal complement of the tangent space, the orthogonal vector $\mathbf{Z}_{\mathbf{X}}^\perp$ is parameterized by two reals $\alpha$ and $\beta$ such that $\mathbf{Z}_{\mathbf{X}}^\perp = \mathbf{X} \mathbf{U}_{\alpha}^{\beta}$. Lastly, using the fact that the tangent vector $\mathbf{Z}_{\mathbf{X}}$ satisfies the equations in \eref{eq:9}, we conclude that the reals $\alpha$ and $\beta$ are the solution to
\begin{align} 
\begin{pmatrix}
|| \mathbf{X} \mathbf{U}_{1}^{0} ||^2_{\mathbf{X}} & \langle \mathbf{X} \mathbf{U}_{0}^{1}, \mathbf{X} \mathbf{U}_{1}^{0} \rangle_{\mathbf{X}} \\
\langle \mathbf{X} \mathbf{U}_{1}^{0}, \mathbf{X} \mathbf{U}_{0}^{1} \rangle_{\mathbf{X}} & || \mathbf{X} \mathbf{U}_{0}^{1} ||^2_{\mathbf{X}}
\end{pmatrix} \begin{pmatrix}
\alpha \\ \beta
\end{pmatrix} = \begin{pmatrix}
\langle \mathbf{Z}, \mathbf{X} \mathbf{U}_{1}^{0} \rangle_{\mathbf{X}} \\
\langle \mathbf{Z}, \mathbf{X} \mathbf{U}_{0}^{1}\rangle_{\mathbf{X}}
\end{pmatrix}.
\label{alpha_beta_eq}
\end{align}
The above linear system of equations admits a unique solution. Indeed, matrix $\mathbf{S}$ defined by
\begin{equation}
 \mathbf{S} = \begin{pmatrix}
|| \mathbf{X} \mathbf{U}_{1}^{0} ||^2_{\mathbf{X}} & \langle \mathbf{X} \mathbf{U}_{0}^{1}, \mathbf{X} \mathbf{U}_{1}^{0} \rangle_{\mathbf{X}} \\
\langle \mathbf{X} \mathbf{U}_{1}^{0}, \mathbf{X} \mathbf{U}_{0}^{1} \rangle_{\mathbf{X}} & || \mathbf{X} \mathbf{U}_{0}^{1} ||^2_{\mathbf{X}}
\end{pmatrix}
\label{S_mat}
\end{equation}
is a {positive definite matrix} with {$\lambda_1^2 + \lambda^2_2 = ||\mathbf{S}||_F^2 < \text{Tr}^2(\mathbf{S}) = (\lambda_1 + \lambda_2)^2$}, i.e., $\lambda_1 \lambda_2 \neq 0$, {where $\lambda_1$ and $\lambda_2$ are the eigenvalues of the matrix $ \mathbf{S}$}. Therefore, the orthogonal projection onto the tangent space is given by
\begin{align} \label{eq:14}
\Pi_{\mathbf{X}}(\mathbf{Z})= \mathbf{Z} - \mathbf{X} \mathbf{U}_{\alpha}^{\beta}
\end{align}
{with $\alpha$ and $\beta$ defined as the solution to the linear system (\ref{alpha_beta_eq})}.

Finally, applying the orthogonal projection $\Pi_{\mathbf{X}}$ to the Euclidean gradient $\nabla_{\mathbf{X}}f$ results in the expression of the Riemannian gradient $\overline{\nabla}_{\mathbf{X}}f$ as follows:
\begin{align} \label{eq:13}
\overline{\nabla}_{\mathbf{X}}f = \nabla_{\mathbf{X}}f - \mathbf{X} \mathbf{U}_{\alpha}^{\beta}
\end{align}
with the reals $\alpha$ and $\beta$ being the solution to the linear system
	\begin{align} 
		&\mathbf{S} \begin{pmatrix}
			\alpha \\ \beta
		\end{pmatrix} =  \begin{pmatrix}
			\langle \nabla_{\mathbf{X}}f, \mathbf{X} \mathbf{U}_{1}^{0} \rangle_{\mathbf{X}} \\
			\langle \nabla_{\mathbf{X}}f, \mathbf{X} \mathbf{U}_{0}^{1}\rangle_{\mathbf{X}}
		\end{pmatrix}. 
		\label{alpha_beta_grad}
	\end{align}
Given the expression of the Riemannian gradient in \eref{eq:13}, the Riemannian Hessian can be computed as the orthogonal projection of the directional derivative of the Riemannian gradient as illustrated in \eref{eq:15}. Let $\mathbf{X} \in \mathcal{M}$ be a vector on the manifold, $\xi_{\mathbf{X}} \in \mathcal{T}_{\mathbf{X}} \mathcal{M}$ a tangent vector, and $f: \mathcal{M} \longrightarrow \mathbb{R}$ a smooth function. The rest of this manuscript uses the shorthand notation $\dot{f}(\mathbf{X})$ to denote the directional derivative $\text{D}(f(\mathbf{X}))[\xi_{\mathbf{X}}]$. Using the previously defined dot notation, the expression of the Riemannian Hessian is provided in the following corollary.
{\begin{corollary}
The Riemannian Hessian for the {isosceles} triangle manifold has the expression
\begin{align} \label{eq:17}
&\overline{\nabla}^2_{\mathbf{X}}f[\xi_{\mathbf{X}}] =  \Pi_{\mathbf{X}}(\nabla^2_{\mathbf{X}}f[\xi_{\mathbf{X}}] - \xi_{\mathbf{X}} \mathbf{U}_{\alpha}^{\beta} - \mathbf{X} \mathbf{U}_{\dot{\alpha}}^{\dot{\beta}}), 
\end{align}
wherein the expression of the orthogonal projection $\Pi_{\mathbf{X}}$ is given in \eref{eq:14} and $\alpha$ and $\beta$ are the solution to \eref{alpha_beta_grad}, wherein $\mathbf{S}$ is given in \eref{S_mat}, and their directional derivatives $\dot{\alpha}$ and $\dot{\beta}$ are the solution to the system 
\begin{align*}
\mathbf{S} \begin{pmatrix} \dot{\alpha} \\ \dot{\beta} \end{pmatrix} = \begin{pmatrix}
\langle \nabla^2_{\mathbf{X}}f[\xi_{\mathbf{X}}], \mathbf{X} \mathbf{U}_{1}^{0} \rangle_{\mathbf{X}} + \langle\nabla_{\mathbf{X}}f, \xi_{\mathbf{X}} \mathbf{U}_{1}^{0} \rangle_{\mathbf{X}} \\
\langle \nabla^2_{\mathbf{X}}f[\xi_{\mathbf{X}}], \mathbf{X} \mathbf{U}_{0}^{1}\rangle_{\mathbf{X}} +  \langle \nabla_{\mathbf{X}}f, \xi_{\mathbf{X}} \mathbf{U}_{0}^{1}\rangle_{\mathbf{X}}
\end{pmatrix} - \dot{\mathbf{S}} \begin{pmatrix} \alpha \\ \beta \end{pmatrix},
\end{align*}
and the directional derivative of $\mathbf{S}$ in the direction $\xi_{\mathbf{X}}$ is
\begin{align*}
\dot{\mathbf{S}} = 
\left(
\begin{array}{c|c}
\multirow{2}{*}{$2\langle \xi_{\mathbf{X}} \mathbf{U}_{1}^{0}, \mathbf{X} \mathbf{U}_{1}^{0} \rangle_{\mathbf{X}}$} & \langle \xi_{\mathbf{X}} \mathbf{U}_{0}^{1}, \mathbf{X} \mathbf{U}_{1}^{0} \rangle_{\mathbf{X}}  \\
& \quad + \langle \mathbf{X} \mathbf{U}_{0}^{1}, \xi_{\mathbf{X}} \mathbf{U}_{1}^{0} \rangle_{\mathbf{X}} \\  
\hline 
\langle \xi_{\mathbf{X}} \mathbf{U}_{1}^{0}, \mathbf{X} \mathbf{U}_{0}^{1} \rangle_{\mathbf{X}}  & \multirow{2}{*}{$2\langle \xi_{\mathbf{X}} \mathbf{U}_{0}^{1},\mathbf{X} \mathbf{U}_{0}^{1}\rangle_{\mathbf{X}}$}\\
\quad+ \langle \mathbf{X} \mathbf{U}_{1}^{0}, \xi_{\mathbf{X}} \mathbf{U}_{0}^{1} \rangle_{\mathbf{X}} & \\
\end{array}
\right).
\end{align*}
\end{corollary}}

\begin{proof}
The proof of this corollary is omitted herein as it follows from a direct computation of the orthogonal projection of the directional derivative of the Riemannian gradient.
\end{proof}

\subsection{Retraction on the {Isosceles} Triangle Manifold}

Designing a computationally efficient retraction is a crucial step in deriving Riemannian optimization algorithms. While it is relatively easy to design functions that are local retractions around $\mathbf{0}_{\mathbf{X}}$, e.g., $\text{R}_{\mathbf{X}}(\xi_{\mathbf{X}}) = \mathbf{X} + \xi_{\mathbf{X}}$, these retractions often result in Riemannian algorithms with poor performance. Indeed, the resulting iterative optimization algorithm would generate smaller and smaller optimization steps, which would ultimately converge before reaching a critical point of the problem. Luckily, for manifolds defined with only equality constraints such as the {isosceles} triangle manifold of interest in this paper, the following theorem allows us to design retractions that are valid for all tangent vectors\cite{absil2009optimization}
\begin{theorem} \label{th2}
Consider an embedded manifold $\mathcal{M}$ in the Euclidean space $\mathcal{E}$ and let $\mathcal{N}$ be an abstract manifold such that dim($\mathcal{M}$) + dim($\mathcal{N}$) = dim($\mathcal{E}$). Let $\mathcal{E}^*$ be an open subset of $\mathcal{E}$ and assume that there is a diffeomorphism function $\phi: \mathcal{M} \times \mathcal{N} \longrightarrow \mathcal{E}^*$, i.e., $\phi$ is a smooth and bijective function with $\phi^{-1}$ also being smooth. Furthermore, assume there exists an element $\mathbf{I} \in \mathcal{N}$ satisfying $\phi(\mathbf{X},\mathbf{I}) = \mathbf{X}, \ \forall \ \mathbf{X} \in \mathcal{M}$. Under the above assumption, the mapping $\text{R}_{\mathbf{X}}(\xi_{\mathbf{X}}) = \pi_1(\phi^{-1}({\mathbf{X}}+\xi_{\mathbf{X}}))$, where $\pi_1(\mathbf{X},\mathbf{Y}) = \mathbf{X}$, defines a retraction on the manifold $\mathcal{M}$ for all tangent vectors $\xi_{\mathbf{X}} \in \mathcal{T}_{\mathbf{X}}\mathcal{M}$ \cite{absil2009optimization}.
\end{theorem}

The rest of this section exploits the result of \thref{th2} to design a computationally efficient retraction. To that end, define $\mathcal{E}^*$ as a subset of $\mathbb{R}^{3 \times 3}${, such that $\mathbf{Z} \in \mathcal{E}^*$ implies that $2 \mathbf{z}_1^T
(\mathbf{z}_2-\mathbf{z}_3) \neq 0$}. It can easily be seen that $\mathcal{E}^*$ is an open subset of $\mathbb{R}^{3 \times 3}$. Furthermore, let $\mathcal{N} = \mathbb{R}^{2}_*$ be the set of $2$-dimensional vectors $\begin{pmatrix} \alpha \\ \beta \end{pmatrix}$ such that {$\alpha > 0$ and $\beta > 0$}. Since dim($\mathcal{N}$) = $2$, the property dim($\mathcal{M}$) + dim($\mathcal{N}$) = dim($\mathcal{E}$) is satisfied. Now define the function
{\begin{align*}
\phi\left(\mathbf{X},\begin{pmatrix} \alpha \\ \beta \end{pmatrix}\right) = \begin{pmatrix}
\alpha\beta \mathbf{x}_1  \\
\beta\mathbf{x}_2  \\
\beta \mathbf{x}_3
\end{pmatrix}.
\end{align*}}

Note that for any $\mathbf{X} \in \mathcal{M}$, we have $\phi\left(\mathbf{X},\begin{pmatrix} 1 \\ 1 \end{pmatrix}\right) = \mathbf{X}$ as mandated by \thref{th2}. In addition, the smoothness of the function $\phi$ directly derives from its definition as it involves only products. Now let $\mathbf{Z}$ be an arbitrary matrix in $\mathcal{E}^*$. From the expression of $\phi$, it can easily be seen that the first term of the inverse $\pi_1(\phi^{-1})$ can be written as
{\begin{align*}
\begin{pmatrix}
\mathbf{x}_1  \\
\mathbf{x}_2  \\
\mathbf{x}_3
\end{pmatrix} = \lambda\begin{pmatrix}
\gamma \mathbf{z}_1  \\
\mathbf{z}_2  \\
\mathbf{z}_3
\end{pmatrix}
\end{align*}}
for some $\lambda$ and $\gamma$ functions of $\mathbf{Z}$. Therefore, the inverse of the first component is smooth. Consider the transformation {$\mathbf{u}_1 = \gamma \mathbf{z}_1$}, $\mathbf{u}_2 = \mathbf{z}_2$, and $\mathbf{u}_3 = \mathbf{z}_3$. It is easy to see that there exists a unique $\gamma$ such that the equality $(\mathbf{u}_1-\mathbf{u}_2)^T(\mathbf{u}_2-\mathbf{u}_3) = -(\mathbf{u}_1-\mathbf{u}_3)^T(\mathbf{u}_2-\mathbf{u}_3)$ is satisfied. The expression of $\gamma$ is given by
{\begin{align*}
\gamma = \cfrac{(\mathbf{z}_2+\mathbf{z}_3)^T(\mathbf{z}_2-\mathbf{z}_3)}{2 \mathbf{z}_1^T (\mathbf{z}_2-\mathbf{z}_3)}.
\end{align*}}
Finally, the point $\mathbf{X} \in \mathcal{M}$ is obtained by scaling the matrix $\mathbf{U}$ by the quantity $\lambda = \sqrt{\cfrac{d^2 \cos(\frac{\pi}{3})}{(\mathbf{u}_1-\mathbf{u}_3)^T(\mathbf{u}_2-\mathbf{u}_3)}}$, i.e., $\mathbf{X} = \lambda \mathbf{U}$ to obtain the manifold characterization 
\begin{align*}
(\mathbf{x}_1-\mathbf{x}_2)^T(\mathbf{x}_2-\mathbf{x}_3) &= - d^2 \cos(\frac{\pi}{3})\\
(\mathbf{x}_1-\mathbf{x}_3)^T(\mathbf{x}_2-\mathbf{x}_3) &= d^2 \cos(\frac{\pi}{3}).
\end{align*}
Since the expressions of $\gamma$ and $\lambda$ are rational functions of the argument $\mathbf{Z}$ without any pole {as $2 \mathbf{z}_1^T (\mathbf{z}_2-\mathbf{z}_3) \neq 0$} and that $\mathbf{X}$ is obtained by a simple multiplication, it can be concluded that $\phi^{-1}$ is smooth which gives that $\phi$ is a diffeomorphism as requested by \thref{th2}. Finally, combining all of the results above and letting $\mathbf{Z} = \mathbf{X}+\xi_{\mathbf{X}}$, this manuscript proposes the following retraction
{\begin{align} \label{eq:16}
\text{R}_{\mathbf{X}}(\xi_{\mathbf{X}}) = \lambda \begin{pmatrix}
\gamma \mathbf{z}_1  \\
\mathbf{z}_2  \\
\mathbf{z}_3
\end{pmatrix}
\end{align} 
with 
{\begin{align}
\lambda &= \sqrt{\cfrac{d^2 \cos(\frac{\pi}{3})}{\left(\gamma \mathbf{z}_1-\mathbf{z}_3\right)^T(\mathbf{z}_2-\mathbf{z}_3)}} \nonumber \\
\nonumber \\
\gamma &= \cfrac{(\mathbf{z}_2+\mathbf{z}_3)^T(\mathbf{z}_2-\mathbf{z}_3)}{2 \mathbf{z}_1^T (\mathbf{z}_2-\mathbf{z}_3)}.
\label{gamma_eq}
\end{align}}
We highlight that the {isosceles} triangle manifold given by (\ref{eq:8}) includes all the points that need to be localized. As an example, the matrix $\mathbf{Y} \notin \mathcal{M}$ with $\mathbf{y}_1^T(\mathbf{y}_2-\mathbf{y}_3) = 0$ is represented by another point $\mathbf{X} \in \mathcal{M}$ with $\mathbf{x}_1 = \mathbf{y}_2$, $\mathbf{x}_2 = \mathbf{y}_3$ and $\mathbf{x}_3 = \mathbf{y}_1$. Therefore, even under the additional inequality constraint, the proposed algorithm can localize all transmitters arranged as an {isosceles} triangle. Finally, without loss of generality we can consider one of the beacons to be located at the origin $\mathbf{b}_j = [0,0,0]$. The distances from this beacon to the points $\mathbf{x}_2$ and $\mathbf{x}_3$ are very similar when $\mathbf{x}_1^T(\mathbf{x}_2-\mathbf{x}_3)=0$. Therefor, whenever the distances $r_{2j} \approx r_{3j}$, we can relabel the {isosceles} triangle vertices such that $\mathbf{x}_1^T(\mathbf{x}_2-\mathbf{x}_3) \neq 0$ and use the proposed algorithm to estimate $\mathbf{x}_1,\mathbf{x}_2$ and $\mathbf{x}_3$.}
\section{High Accuracy $3$D Location Estimation} \label{sec:hig}

This section exploits the previous results to designs a high-accuracy spatial location estimation method using ultrasound waves and the fixed geometry of the transmitters. Given the non-convex nature of the problem, \sref{secsub:1} {presents two types of initialization; a random initialization on the manifold and an improved initialization through the use of a non-linear least squares solver and the Riemannian geometry of the manifold.} \sref{secsub:2} adapts the template of the Riemannian optimization method to the steepest descent algorithm on the {isosceles} triangle manifold. Finally, \sref{sec:sim} combines the initialization obtained with the proposed Riemannian optimization algorithm to efficiently solve the $3$D spatial location problem of interest in this paper.

\subsection{Initialization and Estimation Accuracy} \label{secsub:1}

Incorporating the geometry of the receivers in the optimization problem turns the problem into a non-convex program. {While the performance of the proposed Riemannian optimization algorithm is slightly affected by the choice of the initial point, the performance of some of the benchmark non-convex solvers that we will compare our method against heavily relies on the quality of the initialization. Therefore, we evaluate the performance of the proposed algorithm against the benchmark algorithms using two methods for initialization; a random initialization and an improved initialization.} A random initialization on the manifold can be obtained by generating a random orthonormal matrix $\mathbf{O} \in \mathbb{R}^{3 \times 3}$, i.e., $\mathbf{O}\mathbf{O}^T=\mathbf{I}$, and initializing $\mathbf{X} = d\sqrt{\cos(\frac{\pi}{3})} \mathbf{O}$. However, due to the non-convex nature of the optimization problem, better results can be obtained by using an improved initialization.

This section proposes an improved initialization by solving the localization problem without constraints on the geometry of the transmitters. This can be accomplished using a classical non-linear least squares solver, such as the Gauss-Newton algorithm \cite{wright1999numerical}. Let $\tilde{\mathbf{X}}_0$ be the solution obtained without constraints on the geometry of the transmitters. Such a solution does not necessarily belong to the {isosceles} triangle manifold. Therefore, the second step of deriving an improved initialization is to ``project" the point $\tilde{\mathbf{X}}_0$ to the manifold. This is accomplished by solving the optimization problem
\begin{align} \label{eq:54}
\mathbf{X}_0 = \arg \min_{\mathbf{X} \in \mathcal{M}} ||\mathbf{X}-\tilde{\mathbf{X}}_0||_2^2
\end{align}

The optimization problem in \eref{eq:54} can be efficiently solved using the geometry derived in \sref{sec:the}. Indeed, as pointed out previously, the proposed framework allows us to optimize any objective function over the {isosceles} triangle manifold, including the function $||\mathbf{X}-\tilde{\mathbf{X}}_0||_2^2$. Random initialization on the manifold, as described above, can be used to solve \eref{eq:54}. The steps of the algorithm are omitted herein as they are provided and described in the next subsection.

Note that the initialization strongly depends on the assumptions on the system and the considered loss function. In other words, while the proposed initialization in \eref{eq:54} performs well for the considered $\ell_2$ loss in \eref{eq:1}, it may not be optimal for different objective functions.

\subsection{Optimization Over the {Isosceles} Triangle Manifold} \label{secsub:2}

The algorithm starts by initialization $\mathbf{X}= \mathbf{X}_0 \in \mathcal{M}$. Then, the algorithm iterates between finding a search direction and updating the current position. As stated in \sref{sec:opt}, the search direction is given by $\xi_{\mathbf{X}} = - \cfrac{\overline{\nabla}_{\mathbf{X}} f}{||\overline{\nabla}_{\mathbf{X}} f||_{\mathbf{X}}}$ wherein the Riemannian gradient is computed according to \eref{eq:13}. The step size $t$ is chosen by backtracking to satisfy the following Wolfe conditions\cite{wolfe1969convergence}
\begin{enumerate}
\item $f(\mathbf{X} + t \xi_{\mathbf{X}}) \leq f(\mathbf{X}) + c_1 t \xi_{\mathbf{X}}^T \nabla_{\mathbf{X}} f$
\item $- \xi_{\mathbf{X}}^T \nabla_{\mathbf{X} + t \xi_{\mathbf{X}}} \leq -c_2 \xi_{\mathbf{X}}^T \nabla_{\mathbf{X}} f$,
\end{enumerate}
for some constants $0 < c_1 < c_2 < 1$. The tangent vector $\xi_{\mathbf{X}}$ scaled with the step size $t$ are retracted to the manifold using \eref{eq:16} to update the position $\mathbf{X}$. The process is repeated until convergence, which can be attested by the norm of the Riemannian gradient. The steps of the proposed Riemannian steepest descent algorithm are summarized in \algref{alg3}.
\begin{algorithm}[t!]
\begin{algorithmic}[1]
\REQUIRE Length $d >0$, initialization $\mathbf{X}_0$, a tolerance $\epsilon >0$, and a smooth function $f$.
\STATE Initialize ${\mathbf{X} = \mathbf{X}_0} \in \mathcal{M}$.
\WHILE {$||\overline{\nabla}_{\mathbf{X}} f||_{{\mathbf{X}}} \neq \epsilon$}
\STATE Find $\alpha$ and $\beta$ by solving (\ref{alpha_beta_grad})
\STATE Compute the Riemannian gradient using (\ref{eq:13})
\STATE Set search direction $\xi_{\mathbf{X}} = - \cfrac{\overline{\nabla}_{\mathbf{X}} f}{||\overline{\nabla}_{\mathbf{X}} f||_{\mathbf{X}}}$
\STATE Compute the step size $t$ using backtracking.
\STATE Define $\mathbf{Z} = \mathbf{X}+ t \xi_{\mathbf{X}}$ and compute $\gamma$ using (\ref{gamma_eq})
\STATE {Define the {isosceles} triangle $\mathbf{U}$ by $$[\mathbf{u}_1,\mathbf{u}_2,\mathbf{u}_3] = \left[ \gamma \mathbf{z}_1,\mathbf{z}_2,\mathbf{z}_3\right] $$}
\STATE Scale the sides of $\mathbf{U}$ to obtain a triangle in $\mathcal{M}$ by $$\mathbf{X} = \sqrt{\cfrac{d^2 \cos(\frac{\pi}{3})}{(\mathbf{u}_1-\mathbf{u}_3)^T(\mathbf{u}_2-\mathbf{u}_3)}} \ \mathbf{U}$$
\ENDWHILE
\end{algorithmic}
\caption{Riemannian Steepest Descent on the {Isosceles} Triangle Manifold for High Accuracy Location Estimation}
\label{alg3}
\end{algorithm}

{Under the notion of convexity in the Euclidean space, it is worth noting that unlike other non-convex methods, whose complexity is at least quadratic with respect to the number of variables $n$, all of the steps in our proposed Riemannian method are linear in $n$. This results in an algorithm with linear complexity overall.} Furthermore, recall that Newton's method on the {isosceles} triangle manifold is obtained by choosing the tangent vector that solves $\overline{\nabla}^2_{\mathbf{X}} f[\xi_{\mathbf{X}}] = -\overline{\nabla}_{\mathbf{X}} f$. Since the previous step can be accomplished in $n^2$ operations, our proposed Riemannian Newton's method is quadratic in terms of the number of variables that competes with the complexity of the first-order generic non-convex solvers, e.g., interior points method (IPM). 
\section{Constrained Cram\'er Rao Bound} \label{sec:ccrb}
The Cram\'er Rao bound (CRB) matrix provides a lower bound on the covariance matrix of any unbiased estimator. In some applications, such as the one described in this paper, the parameters that we intend to estimate are constrained. {To address this problem, several versions of the CRB have been derived for constrained parameter estimation \cite{hendriks1991cramer}, \cite{smith2005covariance}. While the approach by Smith in \cite{smith2005covariance} extends the theory of CRB to parameters on manifold, the steps required to derive the constrained CRB are a bit more complicated than the approach in \cite{700921}. In this section, we derive the CRBs under parametric constraints for our setup using the simpler approach based on  \cite{700921}.} To start with, we derive the unconstrained CRBs for our estimation problem. Then, we utilize these bounds in the constrained CRB (CCRB) theorem \cite{700921} to derive the constrained CRB. 
{\subsection{The unconstrained CRB}
	
Let the received signal from the $i^{\textit{th}}$ transmitter at the $j^{\textit{th}}$ receiver be given by Equation (\ref{ddd}). 
Moreover, let the complex envelope of the received signal be $\zeta_{e_{ij}} (t)$ which can be obtained using an IQ demodulator. We obtain the discrete-time version of this envelope by sampling $\zeta_{e_{ij}} (t)$ at a sampling period $T_s$, which gives
\begin{equation}
\zeta_{e_{ij}}[k] = \psi_{ij} s_{e_i}[k-\tau_{ij}]+n_{ij} [k],
\end{equation}
where $s_{e_i}[k]$ is the discrete-time complex envelope of the transmitted signal from the $i^{\textit{th}}$ transmitter, $\tau_{ij}$ is the ToF, $\kappa_{ij}$, normalized by $T_s$ and rounded to the nearest integer, $n_{ij} [k]$ is a discrete-time complex additive Gaussian noise with zero mean and variance $\sigma _{ij}^2$. At a very high sampling rate, we assume that the error due to rounding is negligible. The $n_{ij} [k]$ are assumed to be independent, consequently the received signals $\zeta_{e_{ij}}[k]$ are independent. The time of flight $\tau_{ij}$ is given by 
\begin{equation}
\tau_{ij} = \frac{\|\pmb{\mathrm{x}}_i -\pmb{\mathrm{b}}_j\|_2}{c T_s},
\end{equation}
where $\pmb{\mathrm{x}}_i$ is the unknown 3D location of the $i^{\textit{th}}$ transmitter, $\pmb{\mathrm{b}}_j$ is the known 3D location of the $j^{\textit{th}}$ receiver, and $c$ is the speed of sound. Therefore, the discrete-time complex envelope of the received signal can be re-written as

\begin{equation}
\zeta_{e_{ij}}[k] = \psi_{ij} s_{e_i}[k-\frac{\|\pmb{\mathrm{x}}_i -\pmb{\mathrm{b}}_j\|_2}{c T_s}]+n_{ij} [k].
\end{equation}

In our setup, we have three transmitters and four receivers, hence $M=3$ and $N= 4$. Let the complex envelope of the transmitted signal $s_{e_i}[k]$ be a Zadoff-Chu sequence \cite{1054840} of length $K$ which is given by

\begin{equation}
s_{e_i}[k] = e^{j \phi_i [k]},
 \label{zc}
\end{equation}
where $\phi_i [k]$ is given by
\[
\phi_i [k] = \begin{dcases*}
\frac{\pi R_i}{K} k(k+1) & if $K$ is odd\\
\frac{\pi R_i}{K} k^2 & if $K$ is even, 
\end{dcases*}
\]
where $R_i$ and $K$ are coprime. The attenuation factors $\psi_{ij}$ are deterministic and assumed to be constant over the observation interval; therefore, they will appear as scaling factors on the maximum likelihood estimator.
The probability of a received symbol $\zeta_{e_{ij}}[k]$ conditional on $\pmb{\theta}$, where $\pmb{\theta}=[\pmb{\mathrm{x}}_1,\pmb{\mathrm{x}}_2,\pmb{\mathrm{x}}_3 ]^T$, is given by
\begin{equation}
p(\zeta_{e_{ij}}[k]|\pmb{\theta})= \frac{1}{\pi \sigma_{ij}^2} e^{\frac{-1}{ \sigma_{ij}^2} |\zeta_{e_{ij}}[k] - \psi_{ij} s_{e_i}[k-\frac{\|\pmb{\mathrm{x}}_i -\pmb{\mathrm{b}}_j\|_2}{c T_s}]|^2}.
\end{equation}

The noise samples $n_{ij} [k]$ are independent, hence the probability of the received sequence can be expressed as
\vspace{0.5cm}
\begin{equation}
p(\pmb{\zeta}_{e_{ij}}|\pmb{\theta})=\prod_{k=0}^{K-1} p(\zeta_{e_{ij}}[k]|\pmb{\theta}).
\end{equation}

By taking the natural logarithm of this probability distribution function and expanding the terms inside the summation, the log-likelihood of the received signal can be written as 

{\small \begin{align*}
 \ln p(\pmb{\zeta}_{e_{ij}}|\pmb{\theta}) &= -K \ln (\pi \sigma_{ij}^2) - \frac{1}{ \sigma_{ij}^2} \sum_{k=0}^{K-1} \Big[ |\zeta_{e_{ij}}[k]|^2 + | \psi_{ij} s_{e_i}[k- \\
 & \frac{\|\pmb{\mathrm{x}}_i -\pmb{\mathrm{b}}_j\|_2}{c T_s}]|^2 - \zeta^*_{e_{ij}}[k] \psi_{ij} s_{e_i}[k-\frac{\|\pmb{\mathrm{x}}_i -\pmb{\mathrm{b}}_j\|_2}{c T_s}]  -\\
 & \zeta_{e_{ij}}[k] \psi_{ij} s^*_{e_i}[k-\frac{\|\pmb{\mathrm{x}}_i -\pmb{\mathrm{b}}_j\|_2}{c T_s}] \Big].
\end{align*}}

Since the $\pmb{\zeta}_{e_{ij}}$’s are independent, the log-likelihood function can be expressed as
\begin{equation}
\ln p(\pmb{\zeta}|\pmb{\theta}) = \sum_{i=1}^{M} \sum_{j=1}^{N} \ln p(\pmb{\zeta}_{e_{ij}}|\pmb{\theta}).
\end{equation}
We can write the log-likelihood of the received signal as
{\footnotesize \begin{align*}
\pmb{L}(\pmb{\mathrm{x}}_1,\pmb{\mathrm{x}}_2,\pmb{\mathrm{x}}_3) &= -K\sum_{i=1}^{M} \sum_{j=1}^{N}\ln (\pi \sigma_{ij}^2) - \sum_{i=1}^{M} \sum_{j=1}^{N}  \frac{1}{ \sigma_{ij}^2} \Bigg[ \sum_{k=0}^{K-1} \Big[ |\zeta_{e_{ij}}[k]|^2 + \\
&  |\psi_{ij} s_{e_i}[k-\frac{\|\pmb{\mathrm{x}}_i -\pmb{\mathrm{b}}_j\|_2}{c T_s}]|^2 - \zeta^*_{e_{ij}}[k]  \psi_{ij} s_{e_i}[k-\frac{\|\pmb{\mathrm{x}}_i -\pmb{\mathrm{b}}_j\|_2}{c T_s}]  \\
& - \zeta_{e_{ij}}[k] \psi_{ij} s^*_{e_i}[k-\frac{\|\pmb{\mathrm{x}}_i -\pmb{\mathrm{b}}_j\|_2}{c T_s}] \Big] \Bigg].
\end{align*}}
 Substitute $s_{e_i}[k]=e^{j\phi_i [k]}$ to obtain
{\footnotesize \begin{align*}
\pmb{L}(\pmb{\mathrm{x}}_1,\pmb{\mathrm{x}}_2,\pmb{\mathrm{x}}_3) &= -K\sum_{i=1}^{M} \sum_{j=1}^{N} \big[ \ln (\pi \sigma_{ij}^2) - \frac{\psi_{ij}^2}{\sigma_{ij}^2} \big] - \sum_{i=1}^{M} \sum_{j=1}^{N}  \frac{1}{ \sigma_{ij}^2} \Bigg[ \sum_{k=0}^{K-1}\\
& \Big[ |\zeta_{e_{ij}}[k]|^2 -  \zeta^*_{e_{ij}}[k] \psi_{ij} e^{j\phi_i[k-\frac{\|\pmb{\mathrm{x}}_i -\pmb{\mathrm{b}}_j\|_2}{c T_s}]}  - \zeta_{e_{ij}}[k] \psi_{ij} \\
& e^{-j\phi_i[k-\frac{\|\pmb{\mathrm{x}}_i -\pmb{\mathrm{b}}_j\|_2}{c T_s}]} \Big] \Bigg].
\end{align*}}
The unconstrained Fisher information matrix (FIM) for $\pmb{\mathrm{x}}_1,\pmb{\mathrm{x}}_2,$ and $\pmb{\mathrm{x}}_3$ is given by $J_{ij} = -\mathbb{E}\big[ \frac{\partial^2 \pmb{L}}{\partial \pmb{\mathrm{x}}_i \partial \pmb{\mathrm{x}}_j}\big]$ \cite{kay1993fundamentals}. Since $\mathbb{E}\big[ \frac{\partial^2 \pmb{L}}{\partial \pmb{\mathrm{x}}_i \partial \pmb{\mathrm{x}}_j}\big] = \pmb{0}_{3 \times 3}, \; \forall \; i\neq j$. Thus, the unconstrained FIM can be written as
\[ \pmb{J}= -
\begin{bmatrix}
\mathbb{E}[\pmb{H}^1] & \pmb{0}_{3 \times 3} & \pmb{0}_{3 \times 3} \\
\pmb{0}_{3 \times 3} & \mathbb{E}[\pmb{H}^2] & \pmb{0}_{3 \times 3}\\
\pmb{0}_{3 \times 3} & \pmb{0}_{3 \times 3} & \mathbb{E}[\pmb{H}^3]
\end{bmatrix},
\]
\vspace{0.5 cm}

\begin{figure*}[!b]
	\vspace{-0.5cm}
	\hrule
	\vspace{0.2cm}
{	\begin{equation}
	\pmb{\Psi} = NULL(\pmb{Q}) = \begin{bmatrix}
	-b_{1} b_{2} & -b_{1} b_{3} & a_{1} b_{2}-a_{2} b_{1} & a_{1} b_{3}-a_{3} b_{1} & -b_{1} (a_{1}+c_{1} ) & -a_{1} b_{2}-b_{1} c_{2} & -a_{1} b_{3}-b_{1} c_{3}\\ 
	b^2_{1} & 0 & 0 & 0 & 0 & 0 & 0\\ 0 & b^2_{1} & 0 & 0 & 0 & 0 & 0\\ 
	0 & 0 & -b_{1} b_{2} & -b_{1} b_{3} & b^2_{1} & b_{1} b_{2} & b_{1} b_{3}\\ 
	0 & 0 & b^2_{1} & 0 & 0 & 0 & 0\\ 
	0 & 0 & 0 &  b^2_{1} & 0 & 0 & 0\\ 
	0 & 0 & 0 & 0 & b^2_{1} & 0 & 0\\ 
	0 & 0 & 0 & 0 & 0 & b^2_{1} & 0\\ 
	0 & 0 & 0 & 0 & 0 & 0 & b^2_{1}
	\end{bmatrix}.
	\label{NULL}
	\end{equation}}
\end{figure*}
where $\mathbb{E}[\pmb{H}^i]= \mathbb{E}\big[ \frac{\partial^2 \pmb{L}}{\partial \pmb{\mathrm{x}}_i \partial \pmb{\mathrm{x}}_i}\big] = -\mathbb{E}\big[ (\frac{\partial \pmb{L}}{\partial \pmb{\mathrm{x}}_i})(\frac{\partial \pmb{L}}{\partial \pmb{\mathrm{x}}_i})^T\big]$. With simple algebraic manipulations we obtain the following expression 
{\scriptsize \begin{equation}
\mathbb{E}[\pmb{H}^i] = \sum_{j=1}^{N}  \frac{-2 \psi_{ij}^2}{ \sigma_{ij}^2}  \sum_{k=0}^{K-1} (\nabla_{\pmb{\mathrm{x}}_i} \phi_i[k-\frac{\|\pmb{\mathrm{x}}_i -\pmb{\mathrm{b}}_j\|_2}{c T_s}])(\nabla_{\pmb{\mathrm{x}}_i} \phi_i[k-\frac{\|\pmb{\mathrm{x}}_i -\pmb{\mathrm{b}}_j\|_2}{c T_s}])^T,
\end{equation}}
where $\phi_i[k]$ is given by (\ref{zc}) and $\nabla_{\pmb{\mathrm{x}}_i}$ is the gradient with respect to $\pmb{\mathrm{x}}_i$.
For an even-length sequence, the Hessian matrix can be expressed as follows using sum identities on $k$ and $k^2$
\begin{align*}
\mathbb{E}[\pmb{H}^i] &= \sum_{j=1}^{N} -8(\frac{ \psi_{ij} \pi R_i}{\sigma_{ij} K c T_s})^2 \Bigg[\frac{K}{c^2 T_s^2} - \frac{K(K-1)}{c T_s \|\pmb{\mathrm{x}}_i -\pmb{\mathrm{b}}_j\|_2} + \\
& \frac{K(K-1)(2K-1)}{6 \|\pmb{\mathrm{x}}_i -\pmb{\mathrm{b}}_j\|_2^2}\Bigg](\pmb{\mathrm{x}}_i -\pmb{\mathrm{b}}_j)(\pmb{\mathrm{x}}_i -\pmb{\mathrm{b}}_j)^T.
\end{align*}
Similarly, for an odd-length sequence, the expectation of the Hessian matrix is given by
\begin{align*}
\mathbb{E}[\pmb{H}^i] &= \sum_{j=1}^{N} -2 (\frac{ \psi_{ij} \pi R_i}{ \sigma_{ij} K c T_s})^2 \Bigg[\frac{4K}{c^2 T_s^2} - \frac{4K^2}{c T_s \|\pmb{\mathrm{x}}_i -\pmb{\mathrm{b}}_j\|_2} + \\
& \frac{K(2K-1)(2K+1)}{3 \|\pmb{\mathrm{x}}_i -\pmb{\mathrm{b}}_j\|_2^2}\Bigg](\pmb{\mathrm{x}}_i -\pmb{\mathrm{b}}_j)(\pmb{\mathrm{x}}_i -\pmb{\mathrm{b}}_j)^T.
\end{align*}
If the FIM $\pmb{J}$ is nonsingular, the unconstrained CRBs are given by \cite{kay1993fundamentals}
\begin{eqnarray}
CRB \geq \pmb{J}^{-1}.
\end{eqnarray}
In the next section, we utilize the unconstrained FIM to obtain the constrained CRBs.}
\subsection{The constrained CRB}

This section provides the CRB for estimating the 3D position of the three transmitters under the {isosceles} triangle constraints (\ref{eq:8}), which can be reformulated as 
{\begin{equation}
\pmb{q}(\pmb{x}_1,\pmb{x}_2,\pmb{x}_3) = \begin{bmatrix}
 \|\pmb{\mathrm{x}}_1-\pmb{\mathrm{x}}_2\|^2 - \|\pmb{\mathrm{x}}_3-\pmb{\mathrm{x}}_1\|^2  \\
\|\pmb{\mathrm{x}}_2-\pmb{\mathrm{x}}_3\|^2 - d^2  \\
\end{bmatrix}  = \pmb{0}.
\label{Cmat}
\end{equation}}
The two constraints given by (\ref{Cmat}) are continuously differentiable. Let us denote the vector that we would like to estimate as $\pmb{\theta} = [\pmb{x}_1,\pmb{x}_2,\pmb{x}_3]^T \in \mathbb{R}^{9 \times 1}$. Moreover, let the $2 \times 9$ Jacobian matrix of the constraints be defined as
{\begin{equation}
\pmb{Q}(\pmb{\theta}) = \frac{\partial \pmb{q}}{\partial \pmb{\theta}} = 2\begin{bmatrix}
-(\pmb{\mathrm{x}}_2  - \pmb{\mathrm{x}}_3)^T & \!\!\!\!\!   -(\pmb{\mathrm{x}}_1 - \pmb{\mathrm{x}}_2)^T & \!\!\!\!\! -(\pmb{\mathrm{x}}_3 - \pmb{\mathrm{x}}_1)^T  \\
\pmb{0}^T & \!\!\!\!\! (\pmb{\mathrm{x}}_2 - \pmb{\mathrm{x}}_3)^T & \!\!\!\!\! - (\pmb{\mathrm{x}}_2 - \pmb{\mathrm{x}}_3)^T 
\end{bmatrix}.
\end{equation}}
Since there are no redundant constraints in (\ref{Cmat}), the matrix $\pmb{Q}(\pmb{\theta})$ is full row rank for any given $\pmb{\theta}$. Therefore, there exists a matrix $\pmb{\Psi} \in \mathbb{R}^{9 \times 7}$ whose columns form an orthonormal basis for the null space of $\pmb{Q}(\pmb{\theta})$ \cite{700921}, that is 
\begin{equation}
\pmb{Q}(\pmb{\theta}) \pmb{\Psi} = \pmb{0},
\label{U}
\end{equation}
where $\pmb{\Psi}^T \pmb{\Psi} = \pmb{I}$.
We can re-write the Jacobian matrix of the constraints into the following format
{\footnotesize
\begin{equation}
\pmb{Q}(\pmb{\theta})  = \begin{bmatrix}
-b_1 & -b_2 & -b_3 & -a_1 & -a_2 & -a_3 & -c_1 & -c_2 & -c_3 \\
0 & 0 & 0 & b_1 & b_2 & b_3 & -b_1 & -b_2 & -b_3 \\
\end{bmatrix},
\end{equation}
}
where $(\pmb{\mathrm{x}}_1 - \pmb{\mathrm{x}}_2)^T = [a_1, a_2, a_3]$, $(\pmb{\mathrm{x}}_2 - \pmb{\mathrm{x}}_3)^T = [b_1, b_2, b_3]$ and $(\pmb{\mathrm{x}}_3 - \pmb{\mathrm{x}}_1)^T = [c_1, c_2, c_3]$. The null space is given by (\ref{NULL}).

To derive the constrained CRB, this section utilizes the constrained CRB theorem \cite{700921} which can be written as
\begin{theorem}\label{th:CCRB}
	Let $\hat{\pmb{\theta}}$ be an unbiased estimate of $\pmb{\theta}$ satisfying (\ref{Cmat}) and let $\pmb{\Psi}$ be defined as in (\ref{U}). If $\pmb{\Psi}^T\pmb{J}\pmb{\Psi}$ is nonsingular, then the constrained CRB is given by
	\begin{equation}
	CCRB \geq \pmb{\Psi} (\pmb{\Psi}^T\pmb{J}\pmb{\Psi})^{-1} \pmb{\Psi}^T.
	\label{CCRB}
	\end{equation}
\end{theorem}
	We numerically evaluate the CCRB, given by (\ref{CCRB}), under different SNR scenarios and compare the performance of our algorithm against these bounds in the Results section.

\section{ Results} \label{sec:sim}

This section presents the simulation results to evaluate the proposed algorithm in a noisy environment. The first subsection presents the simulation environment  and parameters. The second subsection evaluates the performance of the proposed algorithm against the benchmark methods, namely the interior point method (IPM) \cite{byrd2000trust,byrd1999interior,waltz2006interior}, the active set algorithm \cite{nocedal2006numerical}, and the sequential quadratic programming (SQP) algorithm \cite{nocedal2006numerical}. Moreover, the paper illustrates the improvement in the location estimation accuracy as compared to the commonly employed trilateration algorithm that utilizes the Gauss-Newton (GN) method \cite{foy1976position}. 

Besides the previously mentioned steepest descent and Newton's algorithms on manifolds, this paper implements the Riemannian version of the trust region
method\cite{absil2009optimization}, \cite{byrd1987trust}. These methods can readily be implemented using the geometry derived in \sref{sec:the}. Indeed, while these algorithms require a vector transport $\mathcal{T}$, the expression of such operator can be obtained by exploiting the linear structure of the embedding space as $\mathcal{T}_{\eta_{\mathbf{X}}}(\xi_{\mathbf{X}}) = \Pi_{\text{R}_{\mathbf{X}}(\eta_{\mathbf{X}})}(\xi_{\mathbf{X}})$ (see Proposition 8.1.2 \cite{AbsMahSep2008}). All methods use the same initial point which is obtained using {one of two methods: (1) random initialization on the manifold, or (2)} the GN-based trilateration method and projecting the point onto the {isosceles} triangle manifold. 

All Riemannian algorithms were implemented using the MATLAB toolbox Manopt \cite{manopt} on an Intel Core i5 (2.7 GHz GHz) computer with 8Gb 2.4 GHz DDR3 RAM. For all simulations, the maximum number of iterations was set to $1000$, the optimality tolerance is set to $10^{-10}$, and the step tolerance is set to $10^{-16}$. {All benchmark methods were implemented using MATLAB built-in solvers, which are computationally efficient.}

\subsection{Simulation Setup}

The size of the room, where the target is located, is given by $4$ m x $4$ m x $3$ m. To evaluate the CRB and CCRB, {the locations of the three transmitters were randomly chosen such that they form an isosceles triangle with base length $d = 10$ cm, unless otherwise indicated. The length of the two equal sides of the triangle is chosen randomly using uniform distribution bounded by $d/2$ and $2d$}. The true ranges from the transmitters to the beacons, denoted by the $\kappa_{ij}$'s, are computed. Three different Zadoff-Chu sequences, with a length of $151$ symbols, are assigned to each of the three transmitters. The received signal consists of a delayed version of the transmitted signal and additive Gaussian noise with zero mean and variance $\sigma_{ij}^2$. The distances between the transmitters and receivers can be estimated using the algorithm in our previous work \cite{alsharif2017zadoff}.

The noisy range estimates $\hat{d}_{ij}$ are utilized in the GN-based trilateration algorithm to obtain initial estimates of the locations of the transmitters. These initial locations are projected onto the {isosceles} triangles manifold by solving \eref{eq:54} to obtain an initial point that belongs to the manifold. For each SNR scenario, $1000$ observations are processed. 

The performance of the proposed algorithm is evaluated by comparing {the root mean square error (RMSE), calculated as the square root of the average mean square error} between the genuine and estimated positions of the transmitters. Furthermore, the computational complexity of the proposed algorithm is compared to that of the considered benchmark algorithms by calculating the overall running time required to obtain an estimate of the target position. 

\subsection{Numerical Results}
\begin{figure}[t]
 	\centering
 	\includegraphics[width=0.9\linewidth]{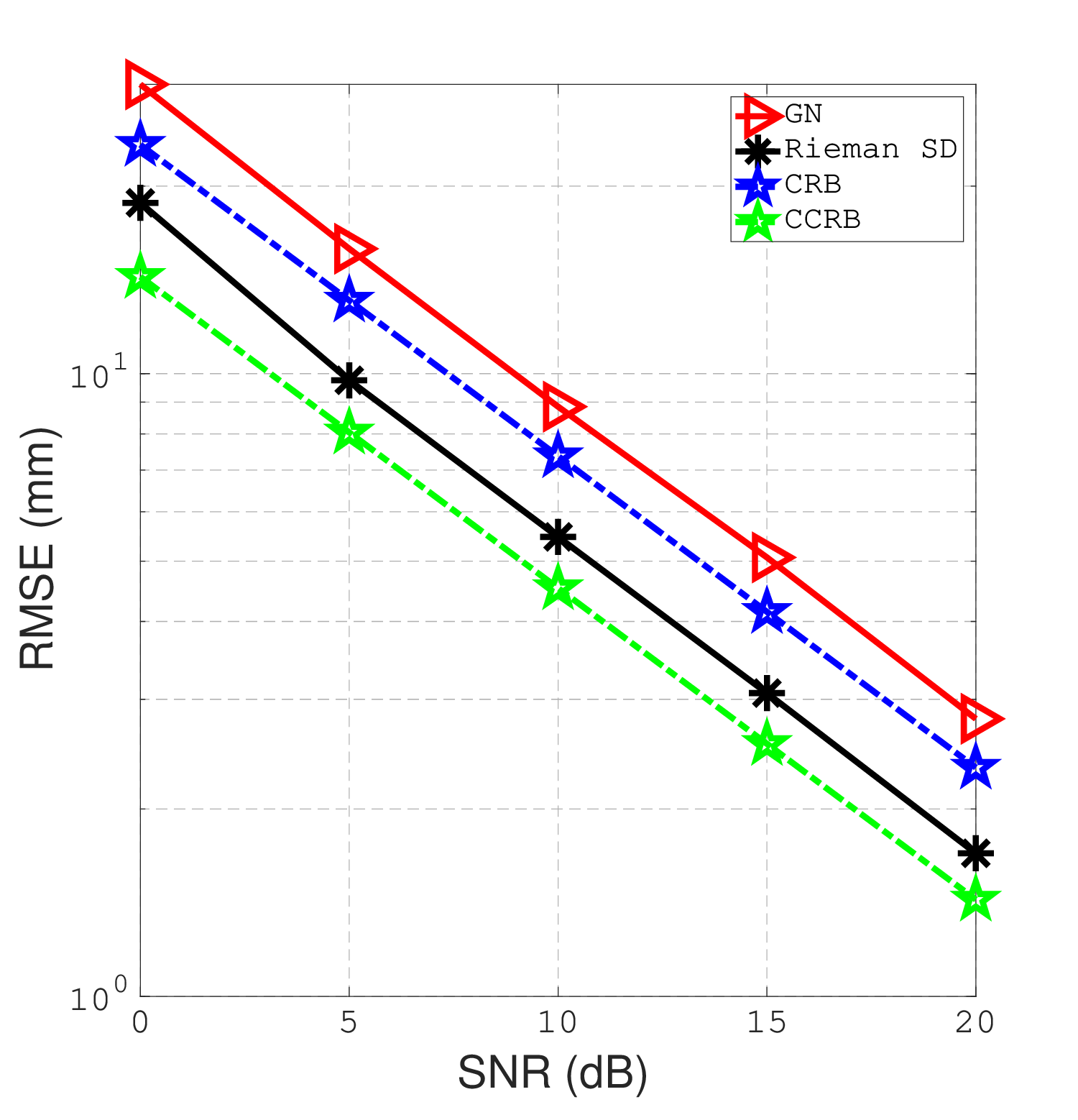}
 	\caption{RMSE vs SNR.} \label{centerError}
 \end{figure}

We illustrate the improvement in localization accuracy obtained through exploiting the {isosceles} triangle geometry, by evaluating the CRB and the constrained CRB. \fref{centerError} plots the RMSE for the GN-based trilateration and the proposed algorithm against the square root of the CRB and the constrained CRB. This plot shows the improvement in the localization accuracy, as the constrained CRB is lower than the unconstrained CRB. {Moreover, the proposed algorithm has a lower RMSE than the unconstrained Gauss-Newton method as expected. Furthermore, we would like to highlight that the accuracy of the localization algorithm is affected by the accuracy of the ranging algorithm. Consequently, the localization algorithm could be further improved by utilizing a more accurate ranging algorithm to estimate the distances between the transmitters and the receivers. {Finally, using the maximum likelihood estimator instead of the nonlinear least squares-based estimator might give an RMSE closer to the CRB and CCRB.} }

\fref{costError} shows the RMSE for the proposed algorithm and the benchmark algorithms {using random initialization. In this simulation, all algorithms iterate until they converge or the number of iterations exceeds a threshold, which was set to $10000$ iterations to guarantee that no method would stop due to being slow.} Both Riemannian algorithms --- the trust region \cite{absil2009optimization}, \cite{byrd1987trust}  and steepest descent-based, given by Algorithm {\ref{alg3}} --- outperform the benchmark algorithms. {Moreover, we noticed that the constrained ASM {and SQP methods do not always converge when using random initialization, hence these two methods give very high RMSEs}. Furthermore, at low SNR values, the proposed algorithms and the {IPM} method give a remarkable improvement as compared to the unconstrained GN-based localization algorithm. The proposed Riemannian-based algorithms maintain their considerable improvement in the localization accuracy at all SNR values. On the contrary, while the performance of the {IPM} algorithm improves as the SNR value increases, these improvements are not as substantial as the improvements provided by the proposed Riemannian algorithms and the unconstrained GN-based algorithm. We recommend that, under random initialization at all SNR values, the proposed Riemannian algorithms should be used to obtain more accurate results than the unconstrained GN-based algorithm and the popular generic non-convex solvers. 
	\begin{figure}[h!]
		\centering
		\begin{subfigure}[h]{0.5\textwidth}
			\centering
			 \includegraphics[width=6.8cm, height = 6.8 cm]{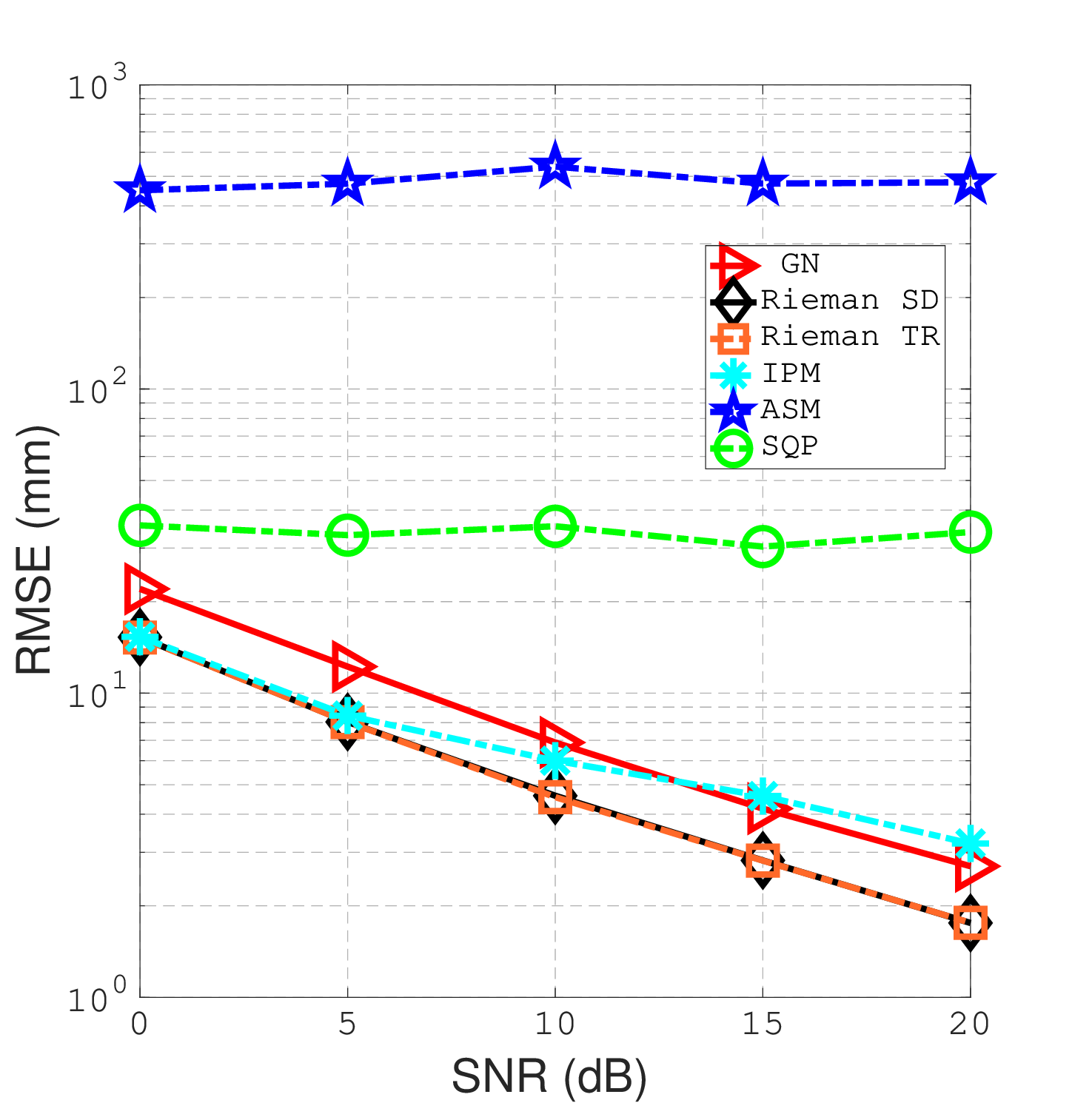}
			\caption{zoomed-out \hspace*{0cm}}
		\end{subfigure}%
		\vspace{-0.1cm}
		\begin{subfigure}[h]{0.5\textwidth}
			\centering
				\includegraphics[width=6.8cm, height = 6.8 cm]{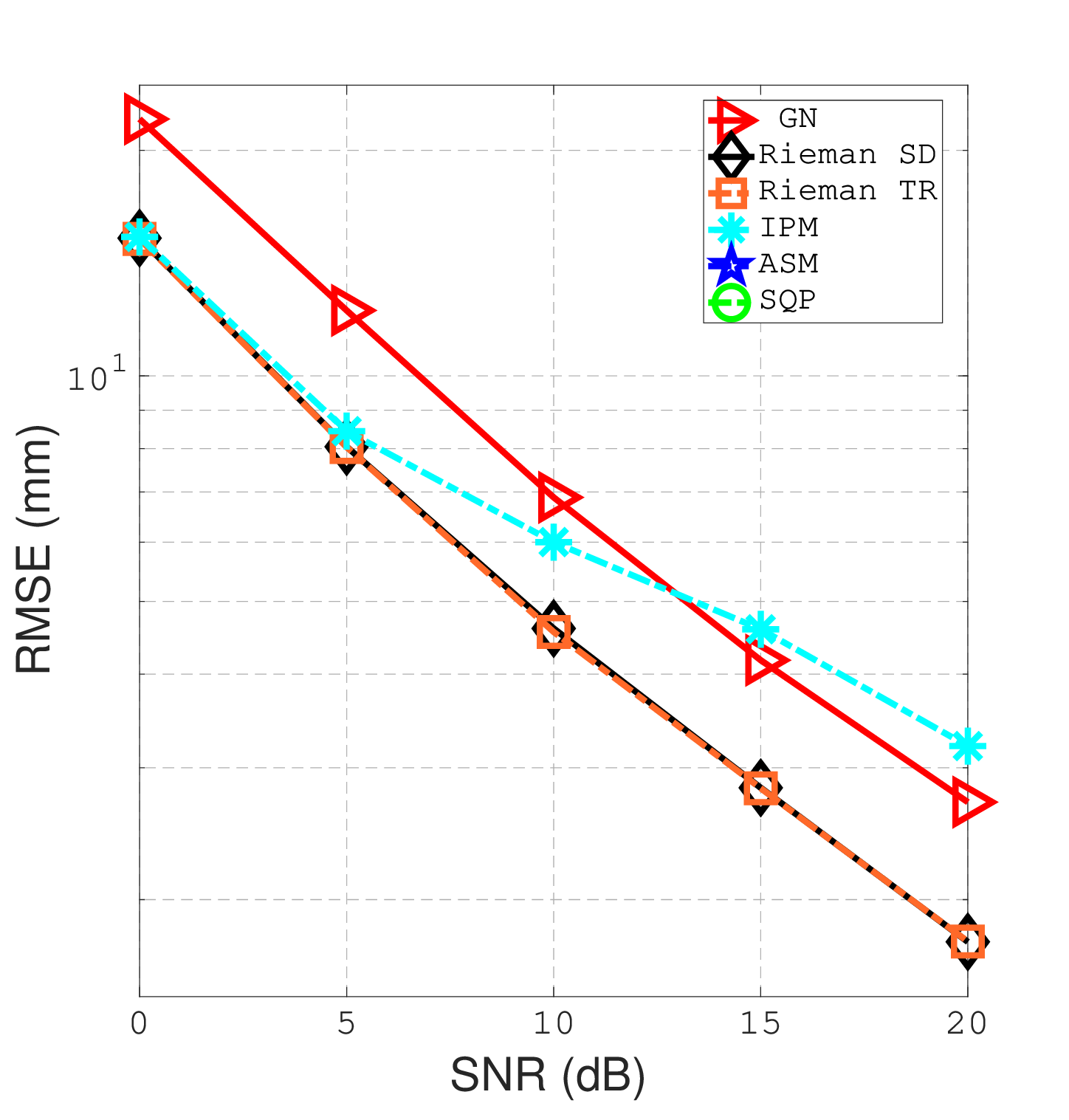}
			\caption{zoomed-in \hspace*{0cm}}
		\end{subfigure}%
		\centering
		\caption{RMSE vs SNR with random initialization.} \label{costError}
	\end{figure}
\begin{figure*}[h!]
	\centering
	\begin{subfigure}[h]{0.5 \textwidth}
		\hfill \includegraphics[width=7cm, height = 6 cm]{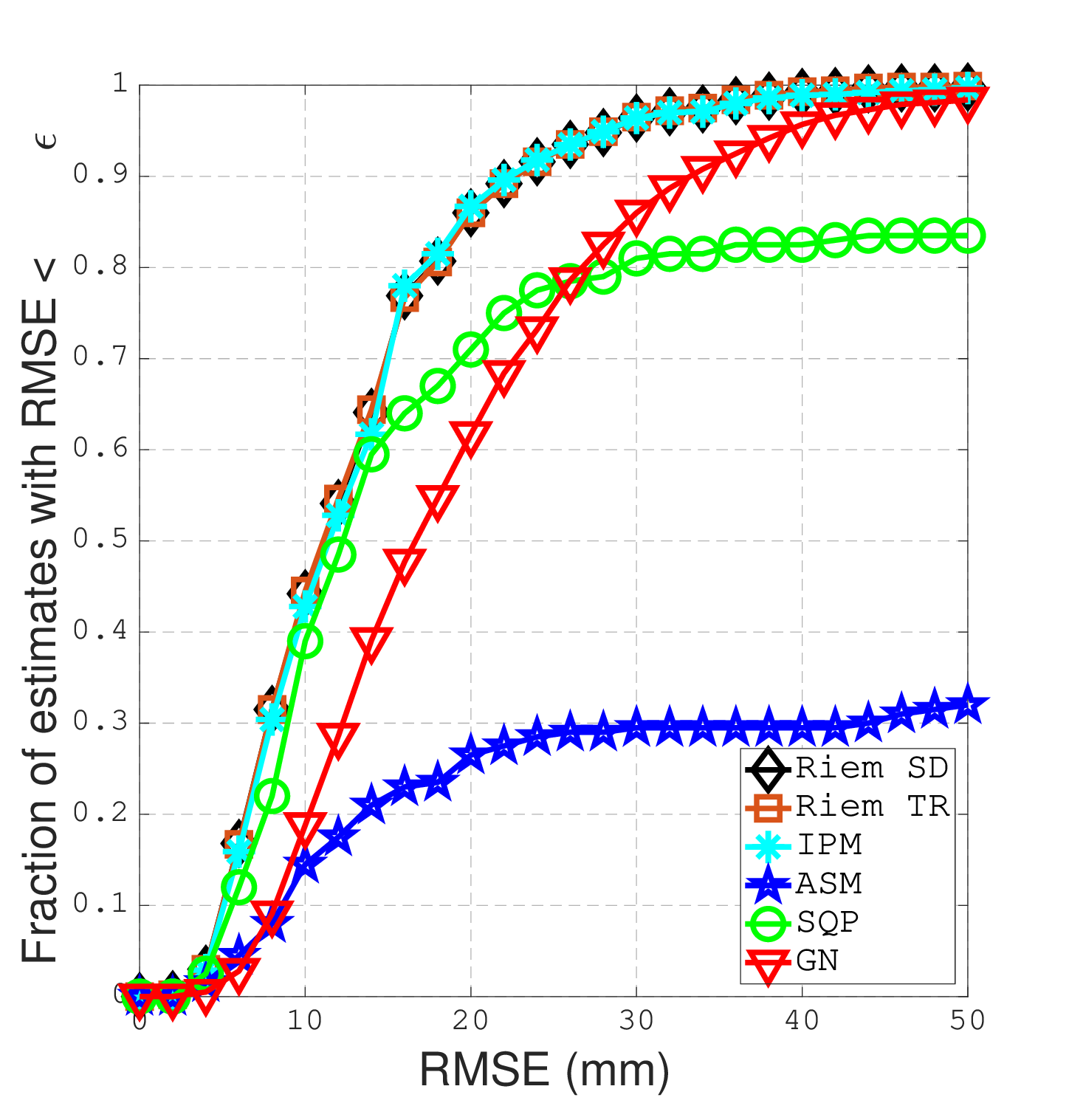}
		\caption{0 dB SNR. \hspace*{-2cm}}
	\end{subfigure}%
	~
	%
	\begin{subfigure}[h]{0.5 \textwidth}
		\includegraphics[width=7cm, height = 6cm]{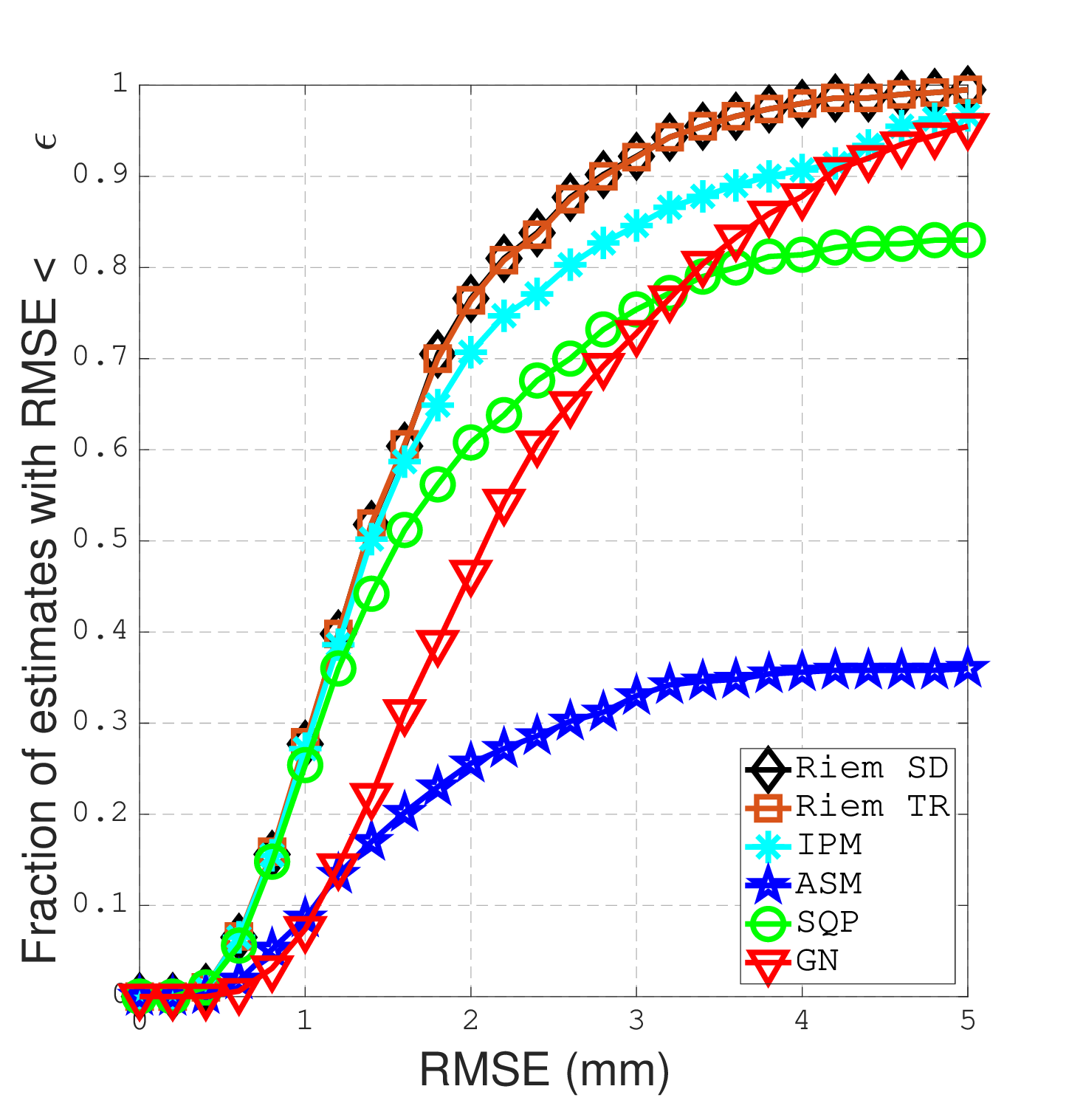}
		\caption{20 dB SNR.\hspace*{1.4cm}}
	\end{subfigure}
	\caption{Cumulative RMSE of the three transmitters for various SNR values with random initialization}
	\label{cumm_RNDM}
\end{figure*}\par
\fref{cumm_RNDM} shows the cumulative error in the estimated locations under two SNR scenarios. The cumulative error plots provide insight into the distribution of the error, hence the effect of outliers on our conclusions will be reduced as compared to conclusions based on the RMSE plots. The proposed algorithm outperforms the benchmark algorithms under all SNR values. In the $0$ dB SNR, over {$98\%$} of the location estimates for the proposed Riemannian algorithm are in less than {$35.4$} mm RMSE. In contrast, less than {$97.5\%$, $82\%$, and $30\%$} of the location estimates for the SQP, IPM and ASM method, respectively, are in less than {$35.4$ mm RMSE}. It is worth noting that while the {SQP} method has a very high RMSE at $0$ dB SNR, as shown in \fref{costError}, its cumulative error is slightly higher than the other benchmark algorithms. This means that the large increase in the RMSE as compared to the other algorithms is due to outliers. 
In the $20$ dB SNR, over $90\%$ of the location estimates for the proposed Riemannian algorithm are in less than {$2.8$} mm RMSE. On the other hand, less than {$83\%$, $73\%$, and $32\%$} of the location estimates for the SQP, IPM and ASM method, respectively, are in less than {$2.8$} mm RMSE.} 

\begin{figure}[h]
	\includegraphics[width=8cm, height = 7 cm]{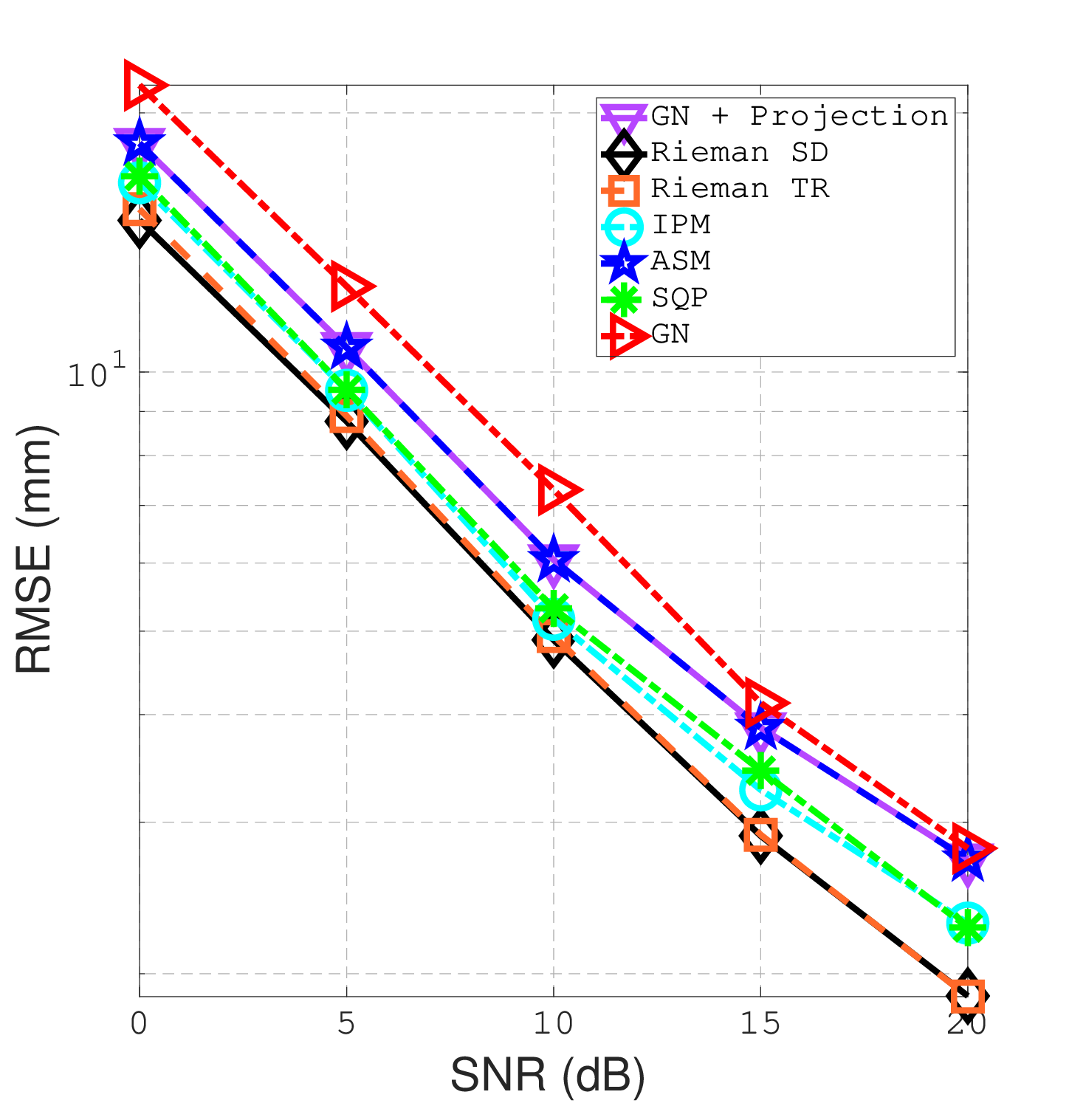}
	\caption{RMSE vs SNR using improved initialization under running time threshold}
	\label{RMSE_improve}
\end{figure}%

\begin{figure*}[h!]
	\centering
	\begin{subfigure}[h]{0.5 \textwidth}
		\hfill \includegraphics[width=7cm, height = 6 cm]{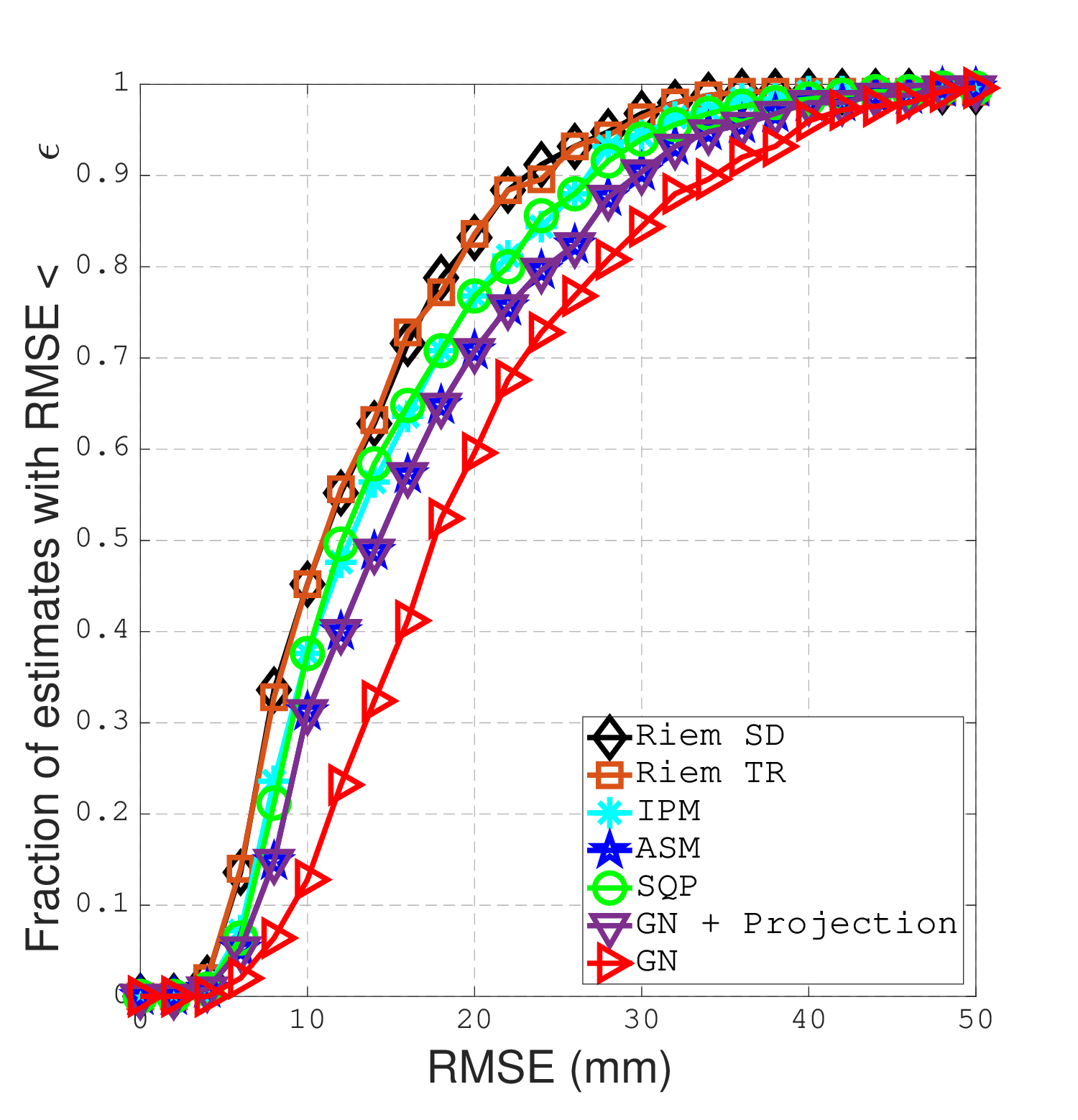}
		\caption{0 dB SNR. \hspace*{-2cm}}
	\end{subfigure}%
	~
	%
	\begin{subfigure}[h]{0.5 \textwidth}
		\includegraphics[width=7cm, height = 6cm]{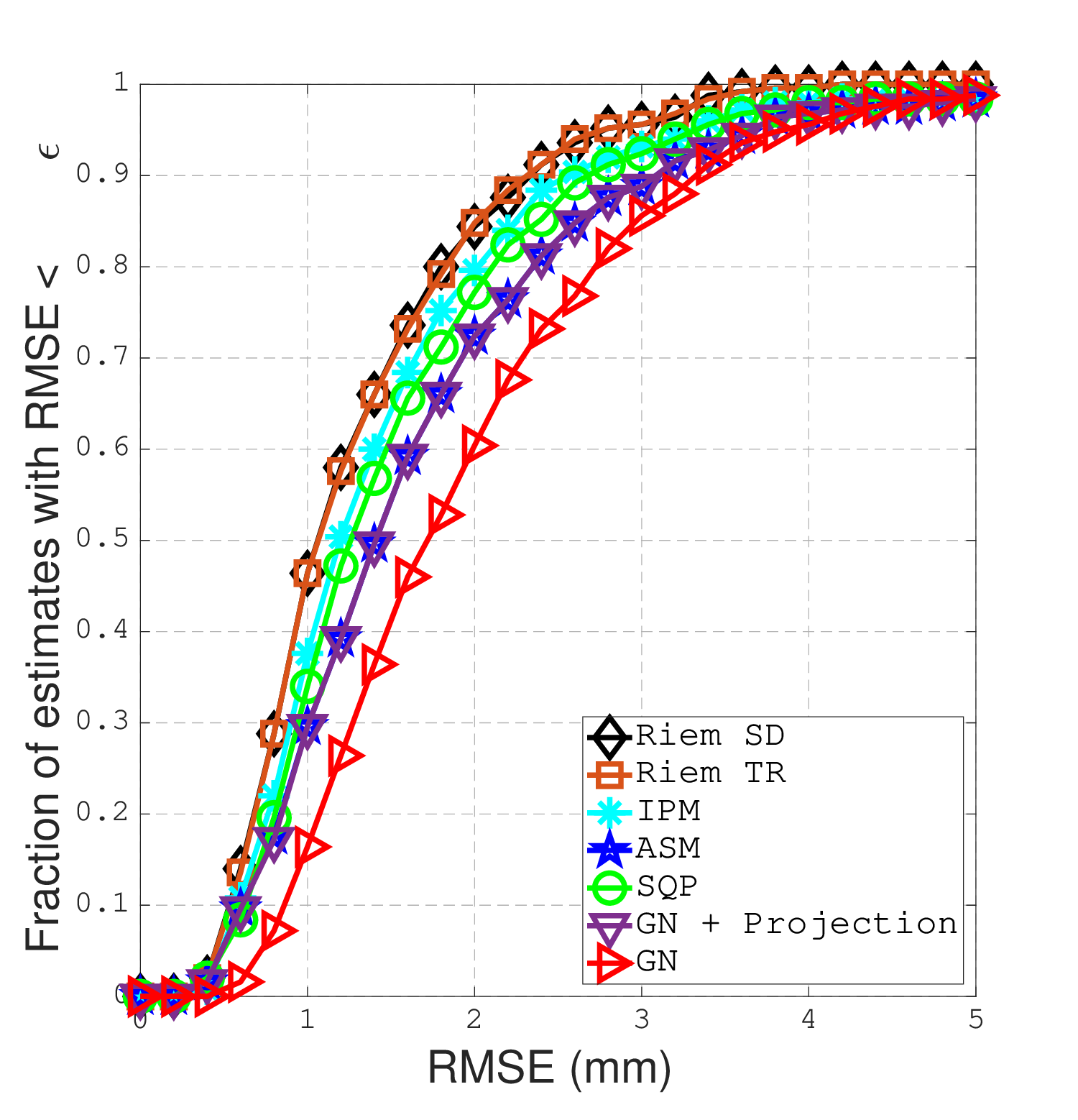}
		\caption{20 dB SNR.\hspace*{1.4cm}}
	\end{subfigure}
	\caption{Cumulative RMSE of the three transmitters for various SNR values with improved initialization}
	\label{cumm}
\end{figure*}

{Since some of the benchmark algorithms are very sensitive to the choice of the initial point, we evaluated the algorithms using an improved initialization, obtained using the algorithm shown in Section VI.A.} {\fref{RMSE_improve} shows the RMSE for the proposed and benchmark algorithms using the improved initialization under a computational time threshold. The computational time threshold is set to make sure that all results are obtained within the same running time. As expected, the improved initial point led to a lower RMSE compared to the unconstrained GN-based estimate. Moreover, the improvement obtained using the ASM method is marginal when compared to the improved initial point. However, we noticed a remarkable improvement when using the IPM and SQP methods compared to the projected GN method. Lastly, both Riemannian-based localization methods outperformed all benchmark algorithms under all SNR scenarios.}

\fref{cumm} shows the cumulative error in the estimated location under two different SNR scenarios {using an improved initialization under a running time threshold}. The proposed algorithm outperforms the benchmark algorithms. {In the $0$ dB SNR, more than $90$ \% of the location estimates for the proposed Riemannian algorithm are in less than {$24 \text{ mm}$} RMSE. On the contrary, less than  $90$ \% of the location estimates for the IPM, Active-set method and SQP method, respectively, are in less than {$27 $ mm, $27 $ mm, and $30 $ mm} RMSE.}
In the $20$ dB SNR, over $90$ \% of the location estimates for the proposed Riemannian algorithm are in less than {$2.3 \text{ mm}$} RMSE. In contrast to the proposed algorithm, less than  $90$ \% of the location estimates for the IPM, Active-set method and SQP method, respectively, are in less than {$2.6$ mm, $3.1 $ mm, and $2.7 $ mm} RMSE. 

\begin{figure}
	\vspace{-1.5cm}
	\centering
	\begin{subfigure}[h!]{0.5 \textwidth}
		\centering
		\includegraphics[width=7cm, height = 7cm]{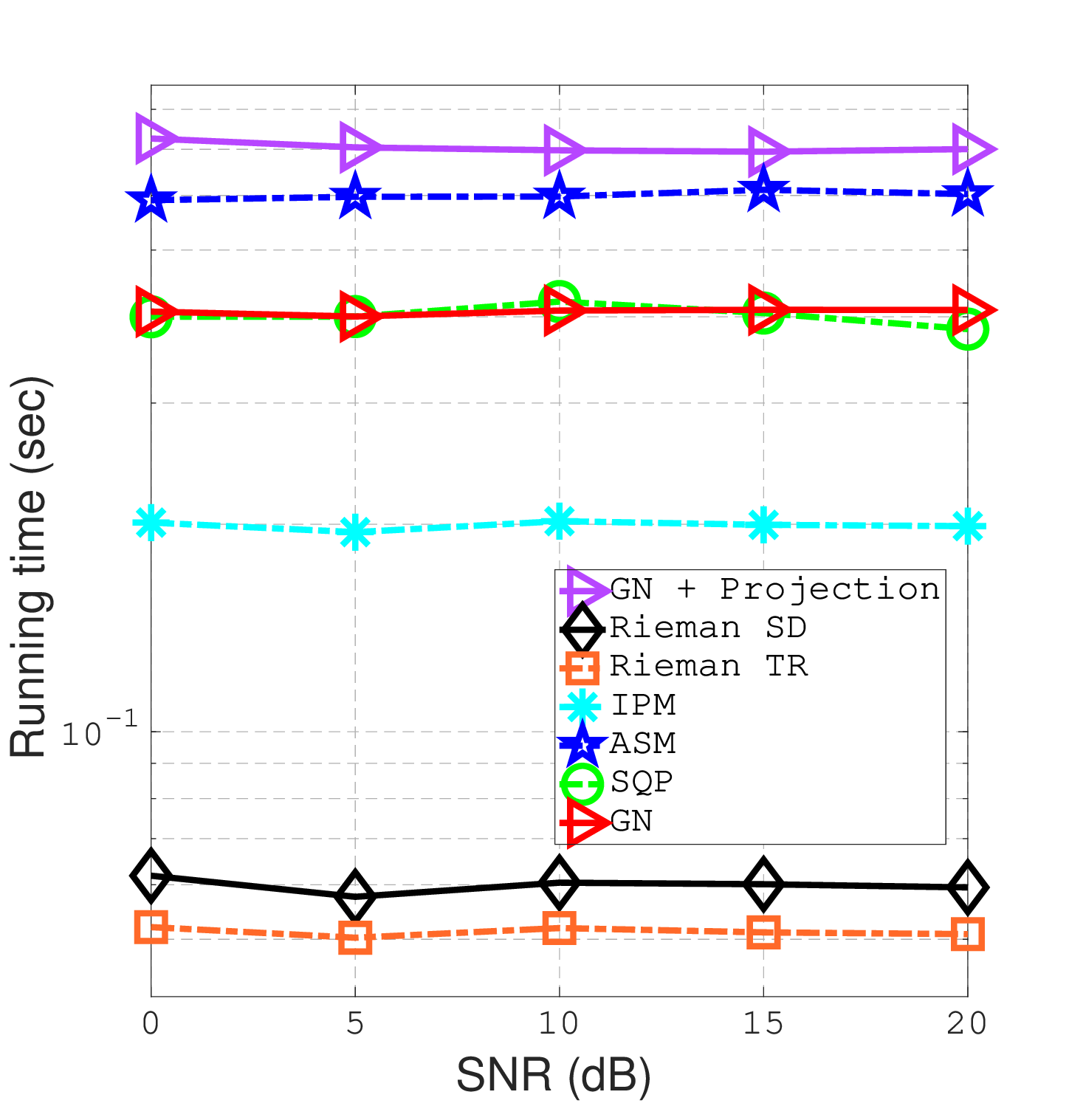}
		\centering
		\caption{Excess running time per estimate. \hspace*{-0.2cm}} \label{time}
	\end{subfigure}%
	
	\begin{subfigure}[h]{0.5 \textwidth}
		\centering
		\includegraphics[width=7cm, height = 7cm]{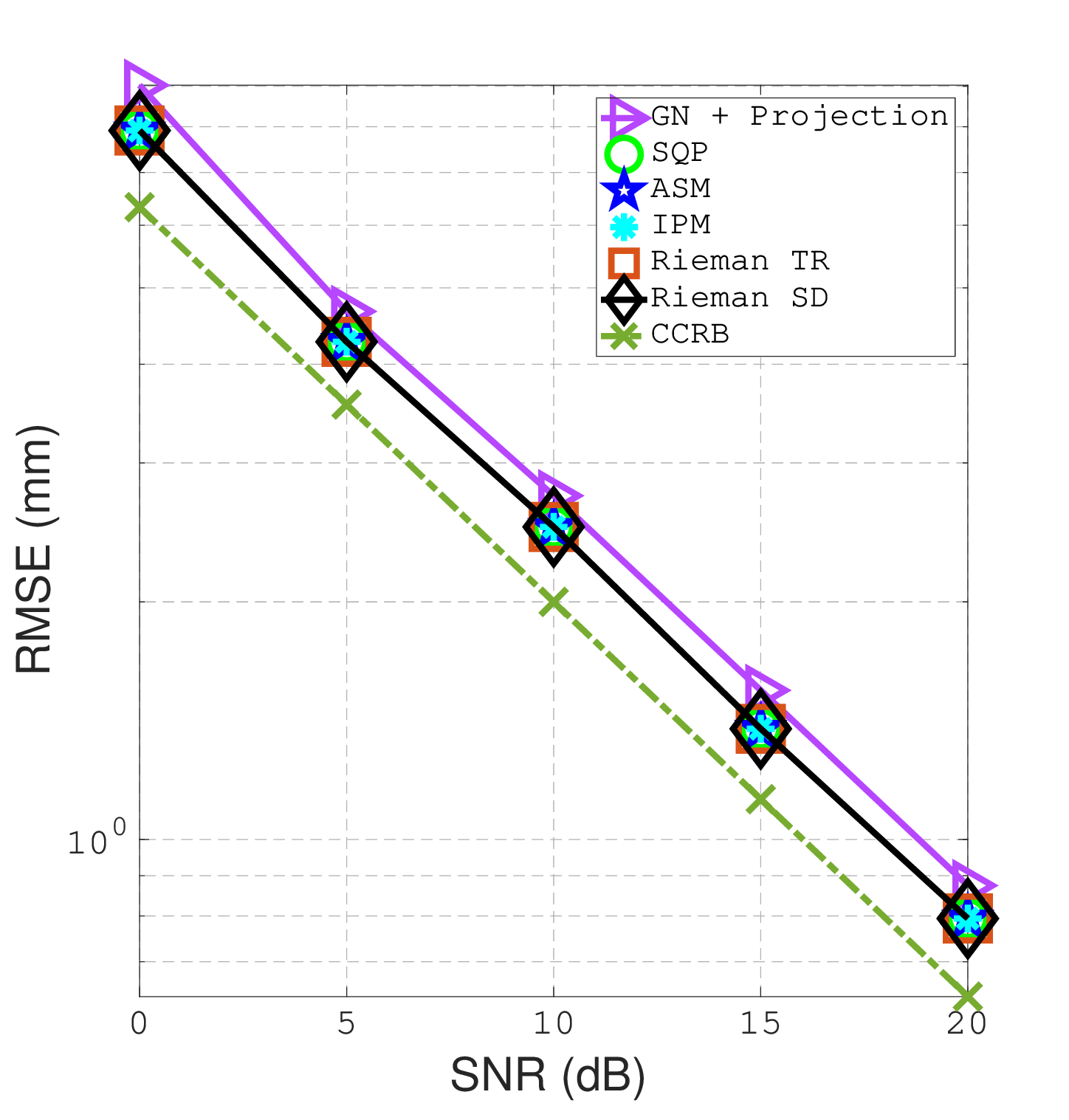}
		\centering
		\caption{MSE vs SNR. \hspace*{-0.2cm}}
	\end{subfigure}
	\caption{Running time and RMSE vs SNR using improved initialization.}
	\label{mse_time_all}
\end{figure}

We compared the computational complexity of all algorithms by calculating the running time required to reach the minimum of the cost function under different SNR values. \fref{mse_time_all} shows both the running time and RMSE for various SNR values {using an improved initialization, obtained by projecting the GN solution onto the manifold}. 
\fref{time} shows that the trust region Riemannian localization algorithm requires much lower running time to reach the minimum compared to all of the benchmark algorithms. {The steepest descent-based Riemannian localization requires more time than the trust region Riemannian method as it has slower convergence rate. {Moreover, Figure 7a shows that the SNR value has negligible influence on the running time.} \fref{mse_time_all}b shows that the constrained localization methods improve the accuracy of the projected GN-based solution. {Finally, Table \ref{table1} and Table \ref{table2} show the running time for all algorithms when using the improved and random initialization, respectively. We would like to highlight that the total time in Table \ref{table1} for each algorithm includes the running time for the improved initialization.} The projected GN method --- used to obtain an improved initial point --- is computationally demanding, which increases the overall computational time for all algorithms. Since the proposed Riemannian algorithm has a good performance under random initialization, unlike the benchmark algorithms, the proposed algorithm is more practical for indoor localization systems.} 

\begin{table}
	\centering
\begin{tabular}{|c|c|c|c|c|}
	\hline
	\multirow{2}{1.5cm}{Algorithm} & \multicolumn{2}{|c|}{Excess Time (sec)} & \multicolumn{2}{|c|}{Total Time (sec)}\\ \cline{2-5}
	& $5$ dB & $20$ dB & $5$ dB & $20$ dB  \\ 
	\hline
		Projected GN  & $-$  & $-$  & $0.705$ & $0.701$    \\ 
	\hline	
	ASM & $0.597$  & $0.603$  & $1.302$ & $1.111$    \\ 
	\hline	 
	SQP & $0.400$  & $0.384$  & $1.105$  & $1.085$   \\ 
	\hline
	IPM  & $0.195$  & $0.199$  & $0.900$ & $0.900$  \\ 
	\hline
	Rieman SD   & $0.058$  & $0.060$  & $0.763$ & $0.761$ \\ 
	\hline
	Rieman TR   & $0.050$  & $0.051$ & $0.755$  & $0.751$ \\ 
	\hline
\end{tabular}
\caption{Running time for all algorithms under different SNR values using improved initialization}
\label{table1}
\end{table}

\begin{table}
	\centering
	\begin{tabular}{|c|c|c|}
		\hline
		\multirow{2}{1.5cm}{Algorithm} & \multicolumn{2}{|c|}{Total Time (sec)} \\ \cline{2-3}
		& $5$ dB & $20$ dB\\ 
		\hline	
		ASM* & $0.709$  & $0.701$   \\ 
		\hline	 
		SQP* & $0.984$  & $0.892$  \\ 
		\hline
		IPM*  & $0.288$  & $0.288$  \\ 
		\hline
		Rieman SD   & $0.146$  & $0.131$  \\ 
		\hline
		Rieman TR   & $0.133$  & $0.116$ \\ 
		\hline
	\end{tabular}
	\caption{Running time for all algorithms under different SNR values using random initialization}
	\label{table2}
\end{table}
\blfootnote{*The benchmark algorithms might converge to a point with a large RMSE under random initialization as shown in Figure \ref{costError}.}
\vspace{-1cm}
\section{Conclusion} \label{sec:con}

This paper exploits the fixed geometry of  three transmitters to improve the accuracy of spatial location estimates using ultrasound waves. The transmitters are assumed to be placed on an {isosceles} triangle, and such information is leveraged in the derivation of a non-convex optimization problem. As the set of feasible solutions admits a Riemannian structure, the manuscript investigates its geometry in order to design efficient optimization algorithms. Simulation results are presented to demonstrate the superiority of the proposed approach in terms of both performance and complexity compared to popular methods from the literature. 

As a future research direction, one could investigate the problem of optimizing the geometry for transmitters to maximize the efficiency of location estimation using ultrasound waves. {In the study of the optimal three-transmitter array geometry, we can evaluate the constrained Cram\'er-Rao bound for triangles with different angles under a fixed triangle area. Moreover, this work could be extended to consider location estimation of a moving target for which the accuracy of the estimated distances degrades.}
\vspace{-0.2cm}
\vspace{-0.2cm}
\bibliographystyle{IEEEtran}
\bibliography{citations}

\end{document}